\documentclass[twocolumn]{aa}
\usepackage{graphicx}
\usepackage{txfonts}
\begin{document}
   \title{Nucleosynthesis in multi-dimensional SNIa explosions}

   \author{C. Travaglio\inst{1,2},
           W. Hillebrandt\inst{3},
           M. Reinecke\inst{4}
          \and
          F.-K. Thielemann\inst{5}
          }

   \offprints{C. Travaglio}

   \institute{Max-Planck Institut f\"ur Astrophysik, Karl-Schwarzschild Strasse 1,
              D-85741 Garching bei M\"unchen, Germany
          \and
              Istituto Nazionale di Astrofisica (INAF) - Osservatorio Astronomico
              di Torino, Via Osservatorio 20, \\ 10025 Pino Torinese (Torino), Italy\\
              \email{travaglio@to.astro.it}
          \and
              Max-Planck Institut f\"ur Astrophysik, Karl-Schwarzschild Strasse 1,
              D-85741 Garching bei M\"unchen, Germany\\
              \email{wfh@mpa-garching.mpg.de}
          \and
              Max-Planck Institut f\"ur Astrophysik, Karl-Schwarzschild Strasse 1,
              D-85741 Garching bei M\"unchen, Germany\\
              \email{martin@mpa-garching.mpg.de}
          \and
              Department of Physics and Astronomy, University of Basel,
              Klingelbergstrasse B2 CH-4056 Basel, Switzerland\\
              \email{fkt@quasar.physik.unibas.ch}
             }

   \date{Received March **, 2004; accepted **, 2004}

   \abstract{
We present the results of nucleosynthesis calculations based on
multi-dimensional (2D and 3D) hydrodynamical simulations of the
thermonuclear burning phase in type Ia supernovae (hereafter
SNIa). The detailed nucleosynthetic yields of our explosion models are
calculated by post-processing the ejecta, using passively advected
tracer particles. The nuclear reaction network employed in computing
the explosive nucleosynthesis contains 383 nuclear species, ranging
from neutrons, protons, and $\alpha$-particles to $^{98}$Mo. Our models
follow the common assumption that SNIa are the explosions of white
dwarfs that have approached the Chandrasekhar mass ($M_{ch}\sim$   
1.39), and are disrupted by thermonuclear fusion of carbon and oxygen.
But in contrast to 1D models which adjust the burning speed to
reproduce lightcurves and spectra, the thermonuclear burning model
applied in this paper does not contain adjustable
parameters. Therefore variations of the explosion energies and 
nucleosynthesis yields are dependent on changes of the initial
conditions only. Here we discuss the nucleosynthetic yields obtained in 2D
and 3D models with two different choices of ignition conditions   
({\it centrally ignited}, in which the spherical initial flame geometry is
perturbated with toroidal rings, and {\it bubbles}, in which
multi-point ignition conditions are simulated), but keeping the
initial composition of the white dwarf unchanged. Constraints imposed on the
hydrodynamical models from nucleosynthesis as well as from
the radial velocity distribution of the elements are discussed in
detail. We show that in our simulations unburned C and O
varies  typically from $\sim$40\% to $\sim$50\% of the total ejected
material. Some of the unburned material remains between the flame
plumes and is concentrated in low
velocity regions at the end of the simulations. This effect is more
pronounced in 2D than in 3D and in models with a small number of
(large) ignition spots. The main differences between all our models and
standard 1D computations are, besides the higher mass fraction of
unburned C and O, the C/O ratio (in our case is typically a factor of 2.5
higher than in 1D computations), and somewhat lower abundances of certain
intermediate mass nuclei such as S, Cl, Ar, K, and Ca, and of $^{56}$Ni.
We also demonstrate that the amount of $^{56}$Ni produced in the explosion
is a very sensitive function of density and temperature. Because 
explosive C and O burning may produce the iron-group
elements and their isotopes in rather different proportions
one can get different $^{56}$Ni-fractions (and thus supernova
luminosities) without changing the kinetic energy of the explosion.
Finally, we show that we need the high resolution multi-point ignition (bubbles)
model to burn most of the material in the center (demonstrating that high
resolution coupled with a large number of ignition spots is crucial to get rid of 
unburned material in a pure deflagration SNIa model).

   }

   \maketitle
%

\section{Introduction}

Type Ia supernovae (SNIa) are known to be stellar explosions with no 
signs of hydrogen and helium in their spectra, but intermediate mass
elements such as Si, S, Ca and Mg near the maximum of their light
curves, and many Fe lines at later times.  In contrast to massive
stars which are the progenitors of Type II supernovae (SNII), SNIa
progenitors are thought to be white dwarfs (WDs) in binary systems (see
Whelan \& Iben~1973, and Hillebrandt \& Niemeyer~2000 for a more
recent review).  In the canonical model the WD, expected to consist
mainly of carbon and oxygen, approaches the Chandrasekhar mass
(M$_{\rm ch}$) through a not yet known mechanism, presumably accretion
from a companion star, and is then disrupted by a thermonuclear
explosion. The declining light is powered by the radioactive decay of
$^{56}$Ni. A strong argument in favor of this scenario is given by the
fact that these explosion models fit quite well the observed light
curves  and spectra (Leibundgut~2001).

Despite the consistency of this general framework with observations
the detailed theory of how SNIa evolve and explode is still subject of
considerable efforts. Over the last three decades, one-dimensional
spherically symmetric models have been used to study the various 
channels that may give rise to a successful SN Ia in terms of the
predicted spectra, light curves, and nucleosynthesis.  Much of this
work was centered on the M$_{\rm ch}$ scenario wherein a C+O white
dwarf accretes H or He from a binary companion (Nomoto, Thielemann, \&
Yokoi~1984) and ignites explosive carbon burning just before it
reaches a critical mass of M$_{\rm ch} \sim 1.39$ M$_\odot$. The      
subsequent explosion produces enough $^{56}$Ni ($\sim 0.6$ M$_\odot$)
and intermediate mass elements to reproduce ``normal'' SN Ia
lightcurves and spectra, provided that the amount of C+O burned at any  
given density is suitably chosen. In 1D models this can be achieved by
parameterizing the thermonuclear flame speed and, if desired, the
density at which a transition to supersonic burning (detonation)
occurs (Khokhlov et al.~1999; Niemeyer~1999). Moreover, some mixing of 
processed matter had to be assumed in order to fit the observed
spectra. Alternative scenarios, including sub-M$_{\rm ch}$ explosions 
and merging white dwarfs (double degenerates), have met with mixed 
success (see e.g. Arnett \& Livne~1994).

More recently it has become possible to perform multi-dimensional 2D
(Livne~1993; Reinecke, Hillebrandt, \& Niemeyer~1999; Lisewski et
al.~2000) and 3D (Reinecke, Hillebrandt, \& Niemeyer~2002a,b; Gamezo
et al.~2003) simulations of an exploding M$_{\rm ch}$-white dwarf. 
The principal difficulty in these models is the fact that the
hydrodynamically unstable and turbulent nuclear flame front develops
structures on much smaller length scales than can numerically be 
resolved. However, this problem can be overcome by ``large eddy  
simulations'', i.e. by employing subgrid-scale models for the
unresolved scales that provide a guess of the effective turbulent 
flame speed on the scale of the computational grid (Niemeyer \&
Hillebrandt~1995a; R\"opke, Niemeyer, \& Hillebrandt~2003). 
{\bf In this flame model, which is well justified 
in the thin-flame regime and is tested in experiments with premixed turbulent
chemical flames, we do not need a detailed prescription of the
nuclear reactions. Instead, the fuel consumption rate is
roughly propartional to the surface area of the flame front
and its normal (turbulent) velocity.} 

Despite of the need of more detailed studies of such subgrid-scale models, 
it is important to stress that the multi-dimensional simulations reach a   
qualitatively different level of predictive power than 1D models. In
particular, the amount of material burned at a given density can not  
longer be fine-tuned but is determined by the fluid motions on the resolved
scales and a particular choice of the subgrid model (Reinecke et 
al.~2002a). Therefore, once the flame model has been fixed numerical
simulations of the thermonuclear explosion of a given white dwarf can 
be done by just choosing the ignition conditions, including the
chemical composition of the WD, the only remaining (physical)
parameter.

The undeniable influence of SNIa explosions on, among others, the
chemical evolution of galaxies makes the quest for solid theoretical
models and nucleosynthetic yields an urgent task. Guided by decades
of modeling and nucleosynthesis calculations in spherically
symmetric models (the protoype being the W7 model of Nomoto et
at.~1984, Iwamoto et al.~1999, Brachwitz et al.~2000), we have 
begun analyzing the detailed nucleosynthetic yields of our explosion 
models by post-processing the ejecta. This has been performed adding 
a ``lagrangian component'' to our Eulerian scheme in the form of tracer 
particles passively advected with the flow in the course of the Eulerian 
calculation. Therefore we record their $T$ and $\rho$ history by 
interpolating the corresponding quantities from the underlying Eulerian 
grid. A similar method of tracer particles in an Eulerian code to calculate 
the nucleosynthesis has been adopted in a previous study of multi-dimensional
nucleosynthesis in core collapse SNe by Nagataki et al.~(1997), and  
more recently in calculations for very massive stars (Maeda et
al.~2002), for core collapse SNe (Travaglio et al.~2004), and for       
Type~Ia SNe first preliminary results have been discussed by Niemeyer et
al.~(2003).

In this paper we present the nuclear yields resulting from several of 
our multi-dimensional supernova simulations, and we compare them to
the standard W7 (Iwamoto et al.~1999, Brachwitz et al.~2000, Thielemann 
et al.~2003) results. 
In Section~2 we summarize the 2D and 3D SNIa calculations, discussing 
different mode of ignition as well as grid resolution of the hydrodynamic 
code. In Section~3 we describe our method to perform nucleosynthesis 
calculations and the nucleosynthesis network adopted for this work.
Finally, in Section~4 we present and discuss our nucleosynthesis results.
In a first step we have performed resolution studies in 2D consisting of
different methods how to distribute the tracer particles, the number of
particles used, and grid resolution of the hydrodynamic code. Although
2D simulations cannot be considered to be realistic, as was discussed
by Reinecke et al.~(2002a), they can serve to guide the more elaborate
3D models. We then discuss the nucleosynthesis resulting from three 3D
models, a centrally ignited model and two models with a few and many 
off-center ignition spots, respectively. It will be shown that the
more realistic ignition conditions (central ignition or many ignition
spots) also predict nucleosynthesis yields closer to the ones observed
in typical SN Ia's.

Concerning nucleosynthesis we will in particular analyze the range and
distribution of $^{56}$Ni masses we are able to produce with our        
present models, and the sensitivity of the
amount and velocity distribution of unburned material ($^{12}$C,
$^{16}$O, $^{22}$Ne) to the ignition conditions of the explosions
which are still a major uncertainty of SN Ia models.


\section{Recent multi-dimensional SNIa calculations}

We have carried out numerical simulations in 2D and 3D, for several 
different ignition conditions, and for different numerical
resolution. Details of these models are given in a series of papers
(Reinecke et al.~1999, 2002a, 2002b). A detailed discussion of these  
models is not the aim of this paper, therefore only a summary of the   
results essential for our nucleosynthesis calculations will be
repeated in this section.

As long as the evolution of the white dwarf before the thermonuclear
runaway remains largely unexplored {\bf (see recent work by Woosley, Wunsch, 
\& Kuhlen~2004 and references therein)}, only very crude constraints can be
put on the flame geometry at the onset of the burning. It appears   
likely that the deflagration sets in at the surface of quietly burning
``hot bubbles''. Nevertheless very little is known about the number,
size and radial distribution of these hot spots. This is a consequence
of the complicated physical processes taking place in the white
dwarf's core during the convective smoldering phase prior to ignition
lasting for $\sim$1000 years.  The long time scales combined with the  
relatively slow convective motions make numerical simulations of this
phase a daunting task which has not been undertaken in its full
complexity so far.  Theoretical considerations and simplified
simulations carried out by Garcia-Senz \& Woosley~(1995) suggest that
fast burning starts on the surface of many small bubbles ($r\le 5$
km), within 100 km of the star's center. Central ignition is another
possible scenario that has been investigated during the last years    
using multidimensional calculations (Niemeyer \& Hillebrandt~1995b;
Khokhlov~1995). 

In this work we present four models: {\it c3\_2d\_512}, a 2D model with
central ignition, and grid size of 512$^2$; {\it c3\_3d\_256}, a 3D
model with central ignition, and grid size of 256$^3$; {\it b5\_3d\_256},
a 3D model with ignition in 5 bubbles, and grid size of 256$^3$; finally, {\it
b30\_3d\_768}, a 3D model with ignition in 30 bubbles, and grid size of
768$^3$. This last one is the model with the highest resolution possible
to evolve with the computer resources available to us, therefore we will
consider it as the 'standard' model for this paper. It achieves a
central resolution of 3.33 km, using a grid consisting of 768$^3$   
zones. In the simulated octant of our model b30\_3d\_768, 30 bubbles  
with a radius of 10 km were distributed randomly.  The bubble
locations were drawn from a Gaussian probability distribution with a  
dispersion of $\sigma$ = 75 km. Bubbles located more than 2.5 $\sigma$
away from the center were rejected. In all models we started the
simulations with a central ignition density of 2.9$\times$10$^9$
gr/cm$^3$. The simulations have been followed up to 1.5 sec. for all the 
models, except for the {\it b30\_3d\_768}. Due to a very high consumption in 
computer time the model {\it b30\_3d\_768} was stopped when no further 
energy was released.

Fig.~1 shows the energy release for the four models mentioned
above. The curves are nearly identical during the first $\sim$0.5 sec
of the simulation.  Owing to the small volume of the bubbles, the 
initial hydrostatic equilibrium is only slightly disturbed. During the
first stages the energy release is therefore lower than in previous   
simulations. Only after the total flame surface has grown considerably
(mostly by deformation of the bubbles), vigorous burning sets in. In
the late explosion phase (after about 0.5 sec) the total energy
differs for the four simulations, and increases moderately with
increasing resolution (see Reinecke et al.~2002a,b for more
detailed discussion). We also note that the 3D centrally ignited and the 
five-bubble models are remarkably similar, even if the centrally ignited
has a relatively faster burning between $\sim$0.5 and $\sim$1 sec.

   \begin{figure}
   \centering
    \includegraphics[width=9cm]{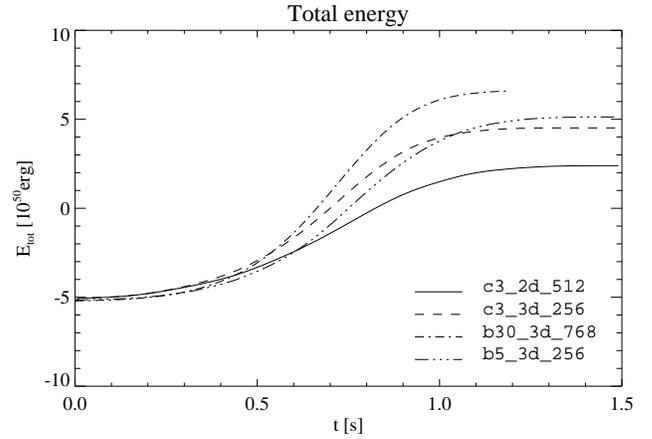}
      \caption{Total energy evolution for the two-dimensional centrally ignited
explosion model ({\it solid line}), for the three-dimensional low-resolution centrally
ignited ({\it dashed line}) explosion model, for the three-dimensional low-resolution 5
bubbles ({\it dotted-dotted-dashed line}), and for the three dimensional high-resolution
30 bubbles ({\it dotted-dashed line}) explosion model.
         \label{fig1}}
   \end{figure}

It must be noted that the five-bubble model is not identical to the model
{\it b5\_3d\_256} presented by Reinecke et al.~(2002b): due to an
oversight during
the simulation setup the initial positions of the burning bubbles are
not the same. As a consequence, the total energy releases of these two
simulations are slightly different.

The initial configuration of the front, as well as snapshots of the   
front evolution at later times are shown in Fig.~2 and Fig.~3 for the
models {\it c3\_3d\_256} and {\it b30\_3d\_768}, respectively.

   \begin{figure*}
   \centering
   \includegraphics[width=6cm]{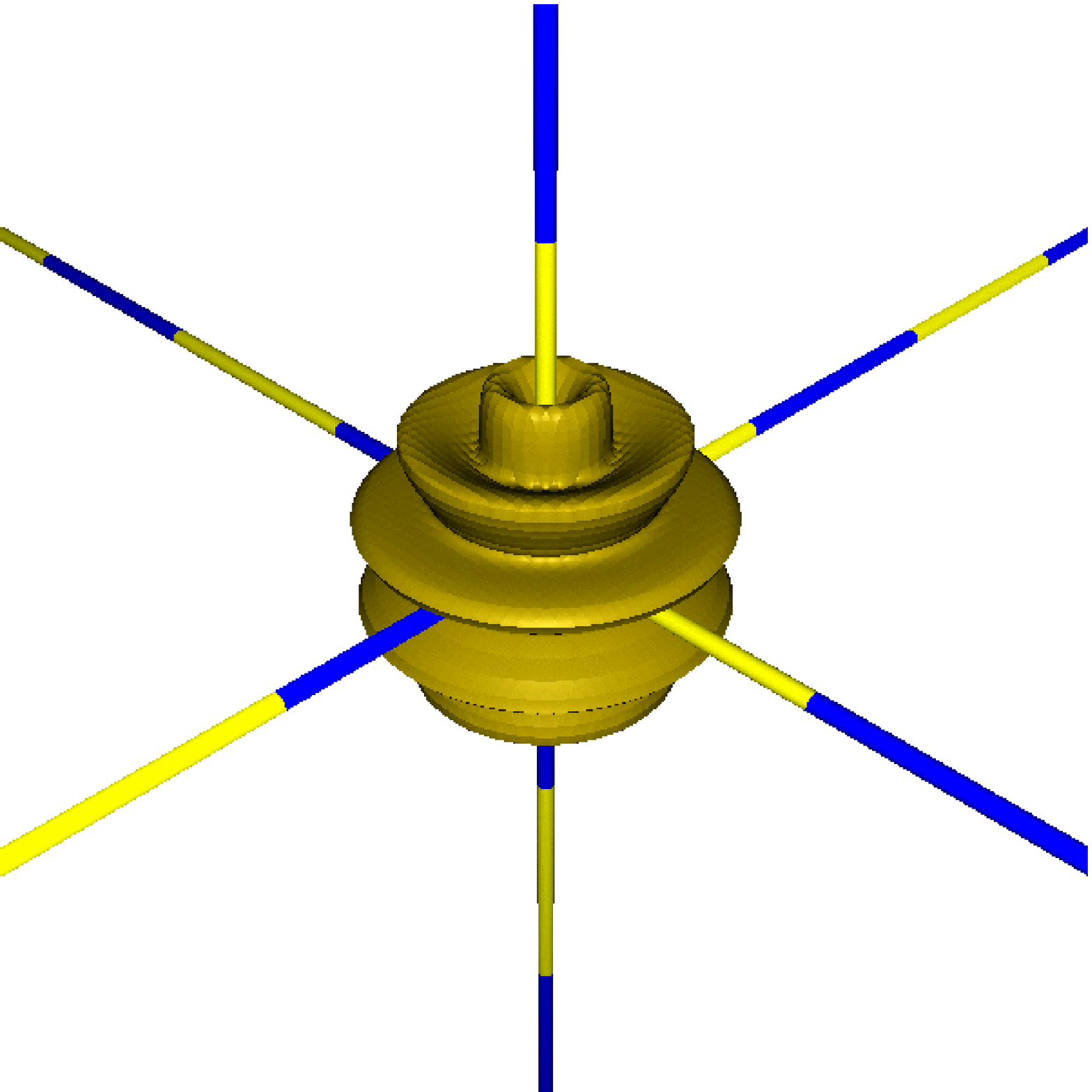}
   \includegraphics[width=6cm]{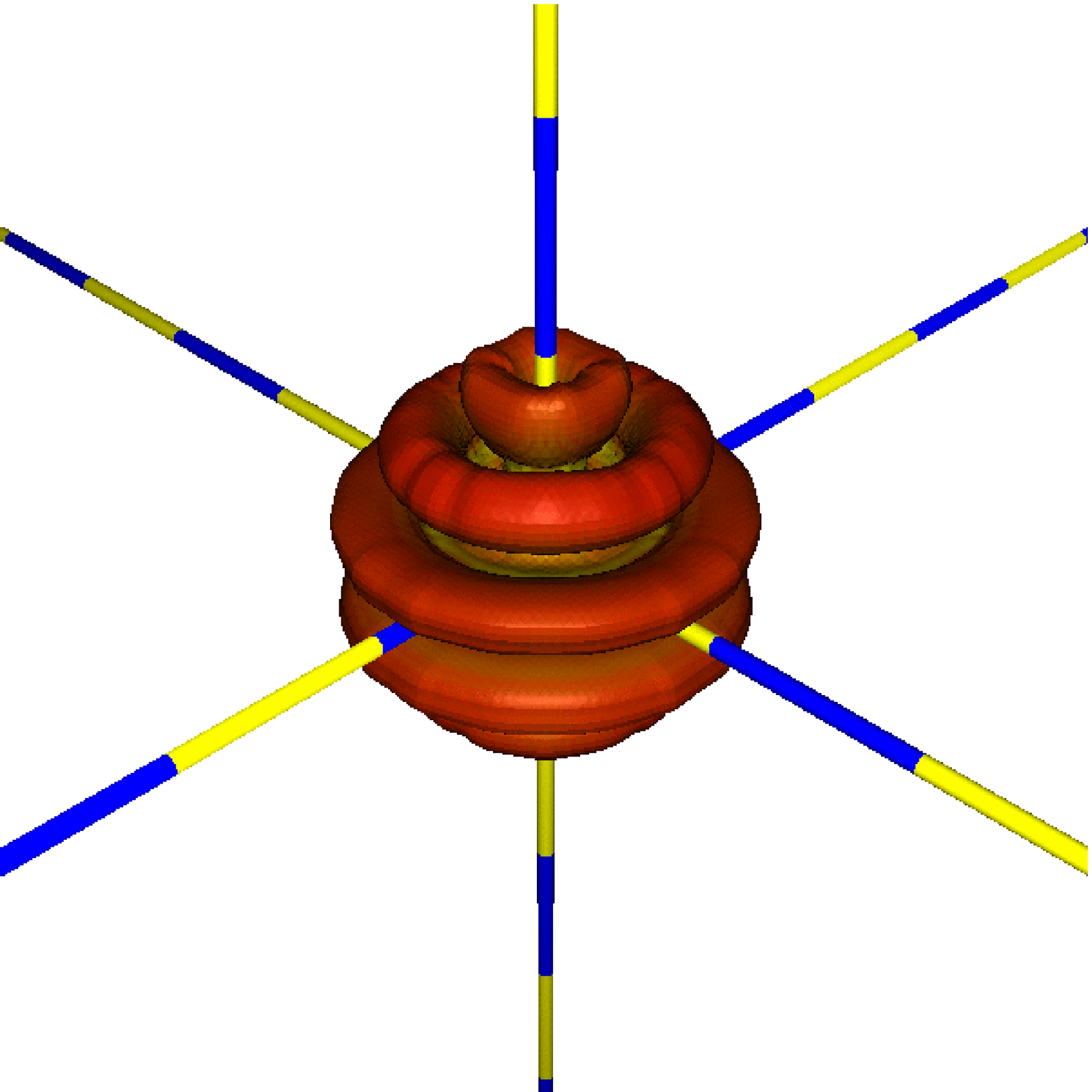}
   \includegraphics[width=6cm]{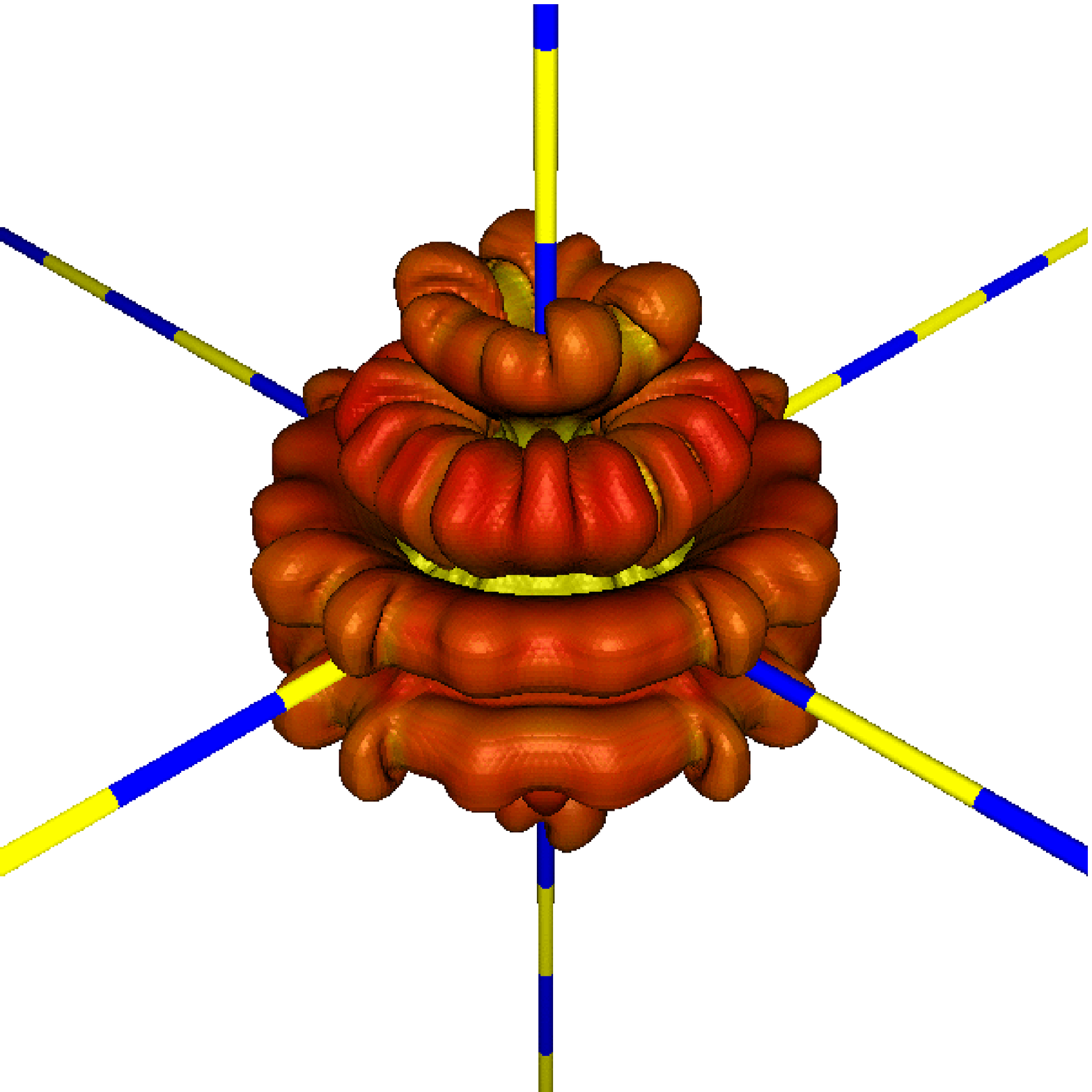}
   \includegraphics[width=6cm]{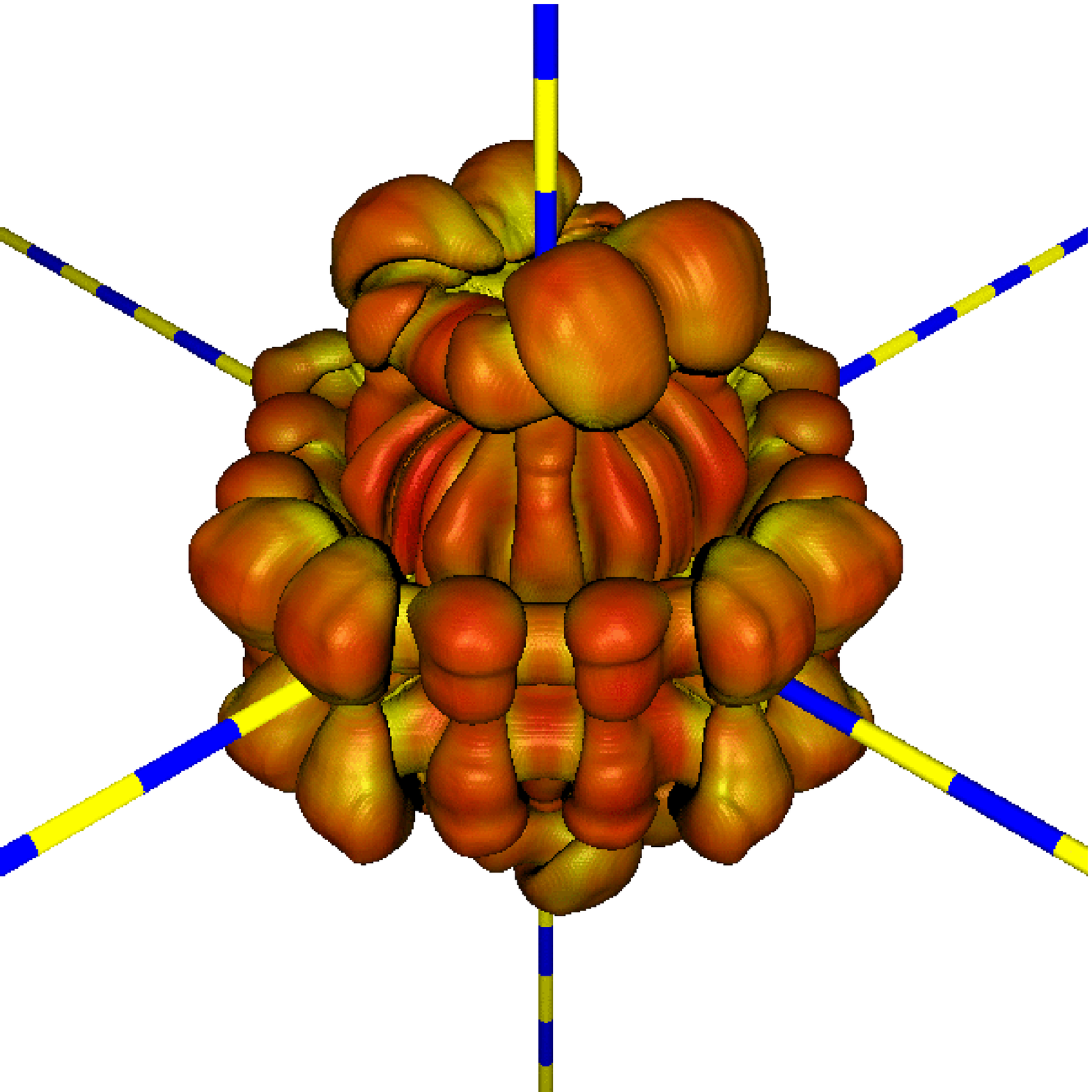}
   \caption{Snapshots of the front evolution for the centrally ignited 
            model {\it
            c3\_3d\_256} at 0 s, 0.2 s, 0.4 s, and 0.6 s.
              \label{fig2}}
    \end{figure*}

   \begin{figure*}
   \centering
   \includegraphics[width=6cm]{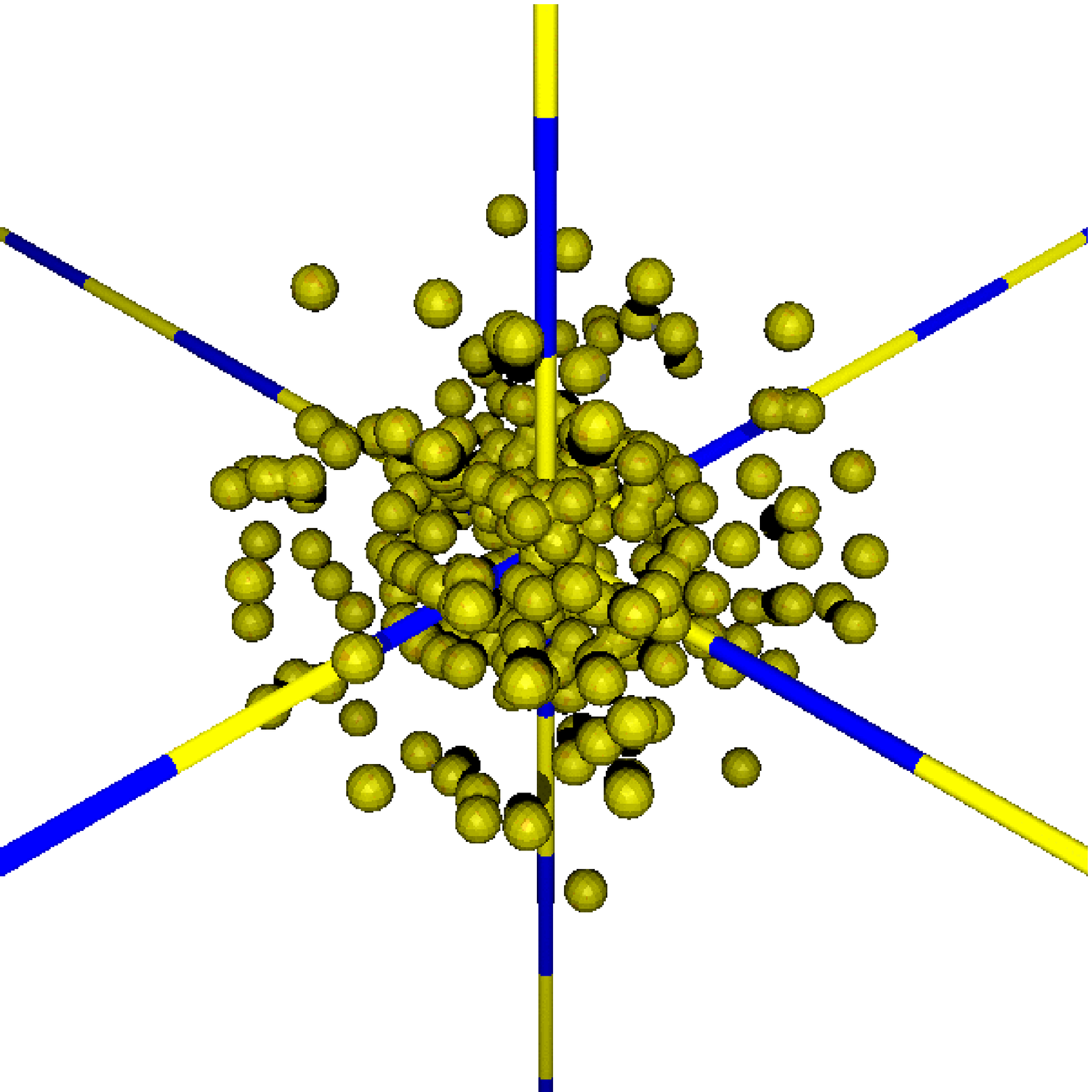}
   \includegraphics[width=6cm]{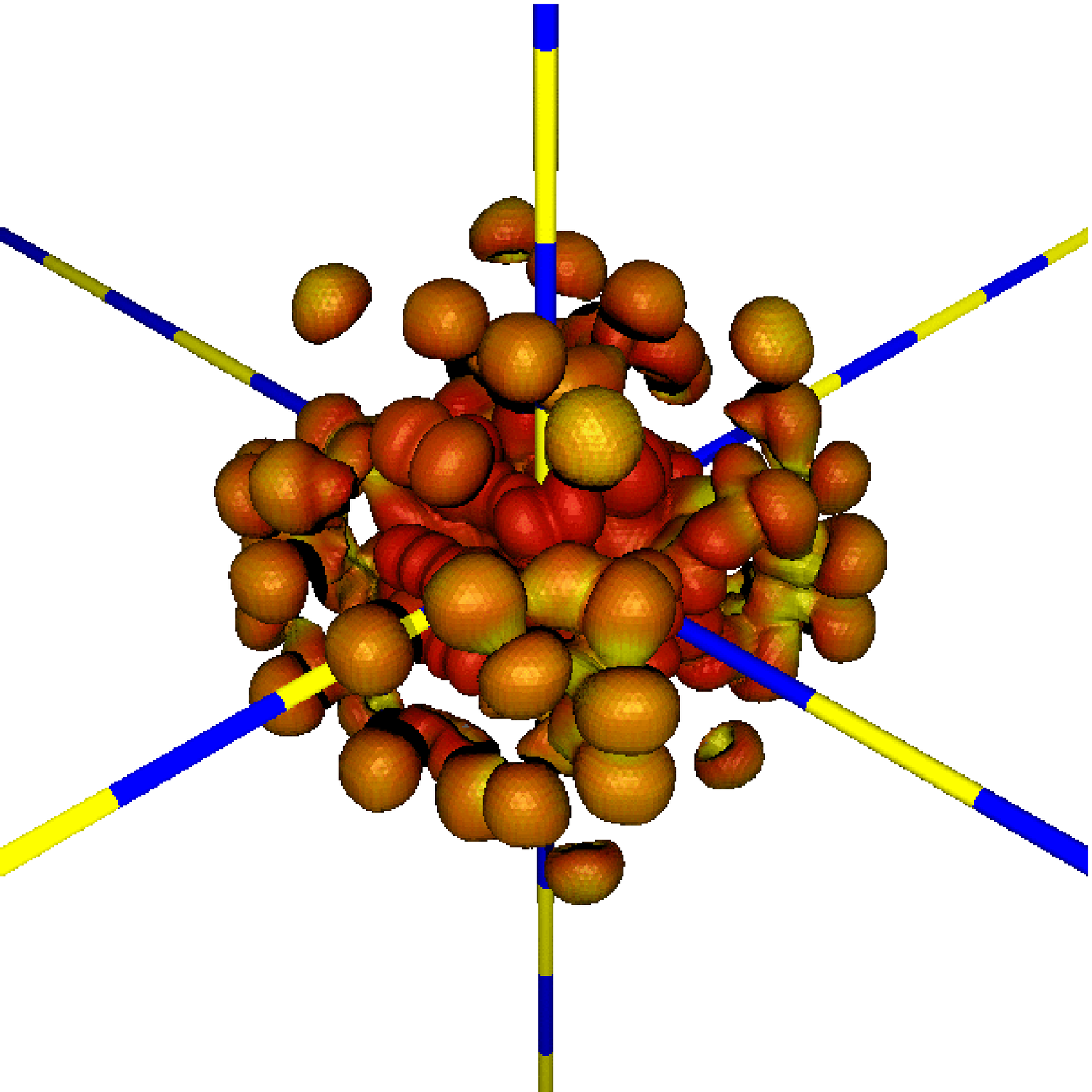}
   \includegraphics[width=6cm]{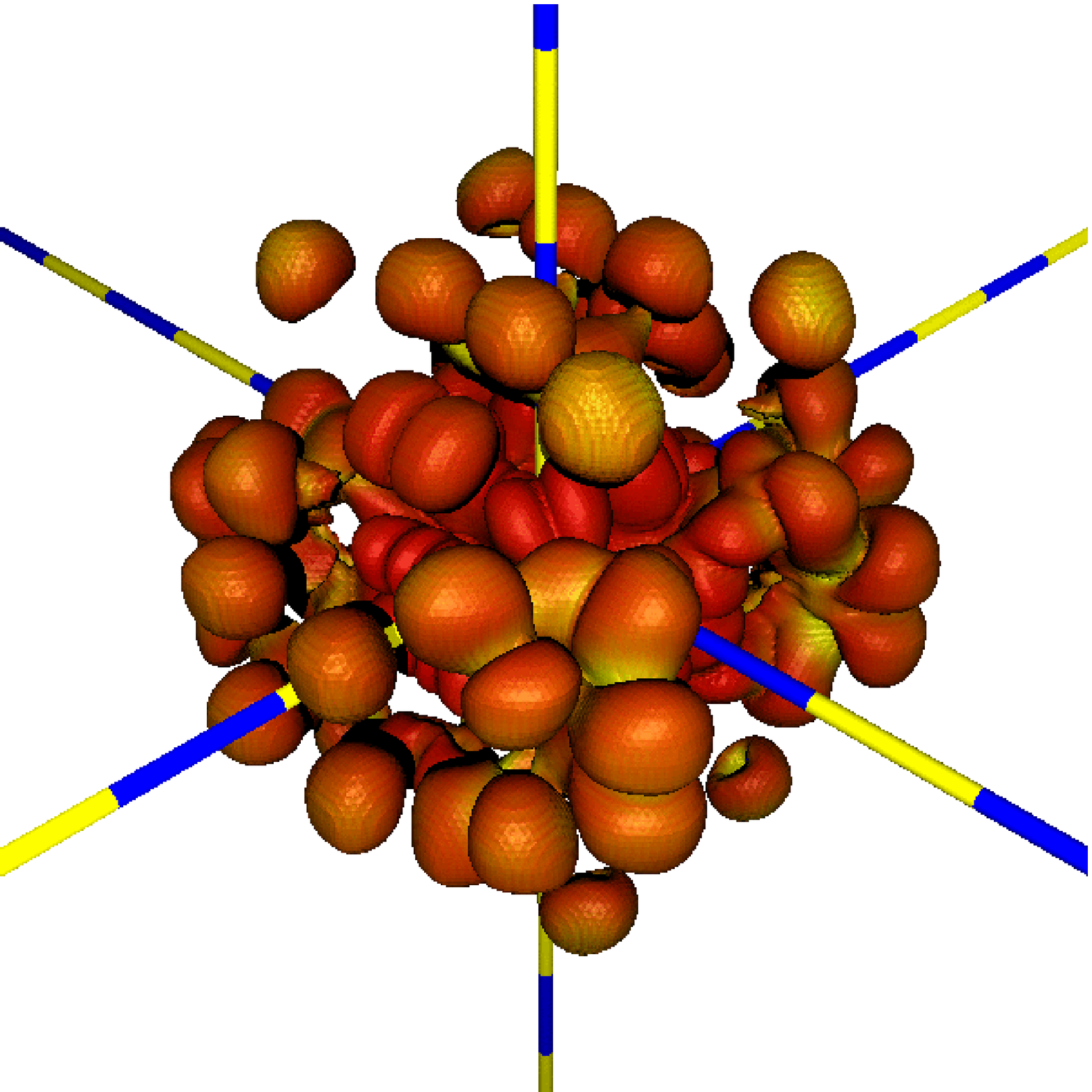}
   \includegraphics[width=6cm]{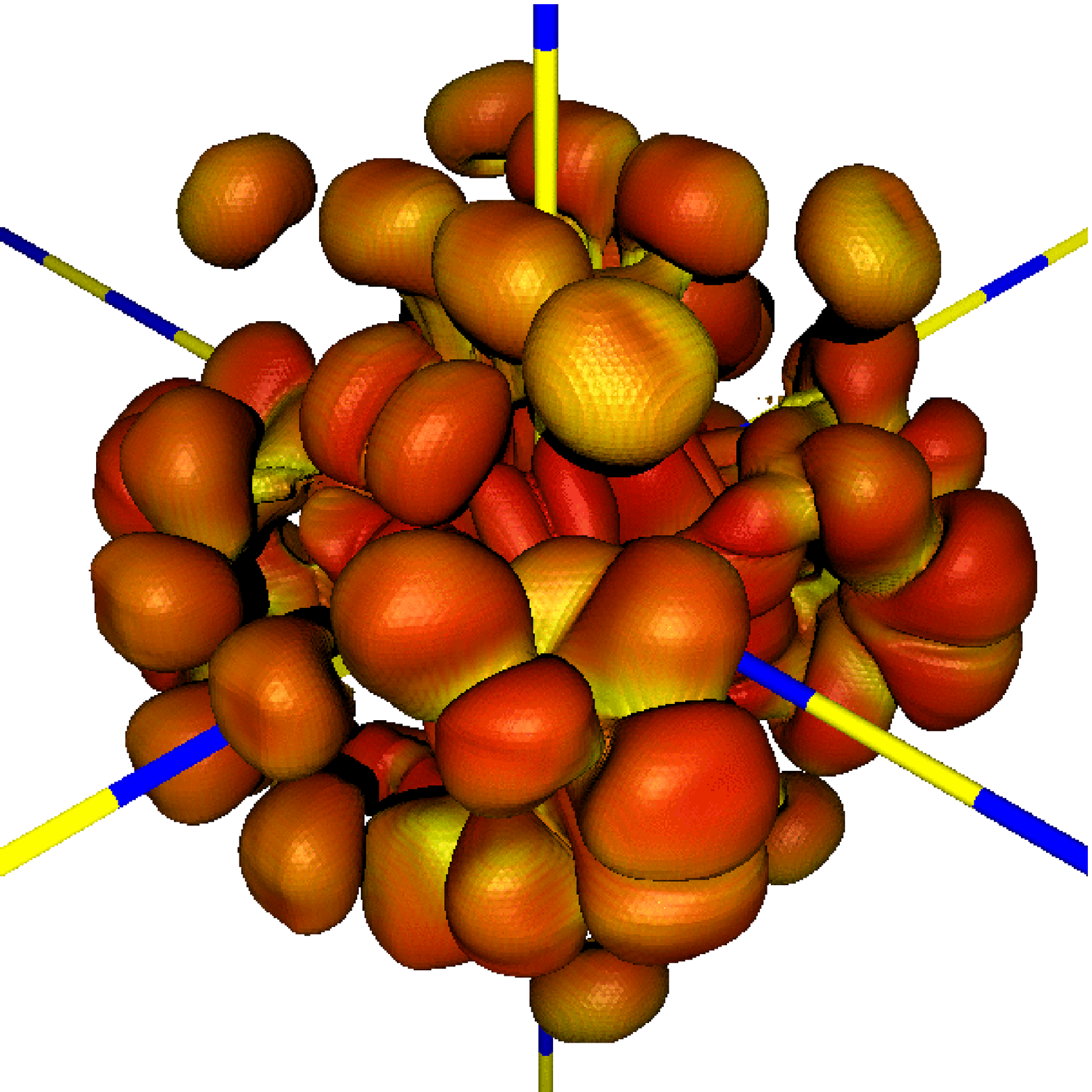}
   \caption{Snapshots of the front evolution for the floating-bubble
            model {\it b30\_3d\_768} at 0 s, 0.1 s, 0.14 s, and 0.2 s. 
            \label{fig3}}
    \end{figure*}

In addition to the total energy release, the mass fraction of unburned
material in the central region of the remnant appears to be a good
criterion for judging the validity of our simulations, because a high
amount of C and O in this region would most likely produce a
characteristic signature in the late-time spectra which has not yet   
been observed. In this respect the results of our earlier calculations
were not very encouraging since the ashes rose towards the surface in large
structures and left nearly pure fuel in the center.
Using many initial bubbles, however, seems to
alleviate this problem insofar as the statistical isotropy of the
initial flame at least delays the development of large-scale turnover
motions. As a consequence, C and O is lower than 20\% in the central
0.2 $M_{\odot}$ after 0.9 s for model {\it b30\_3d\_768}. For this
last model we get $\sim$40\% of the total mass stays unburned
(we define unburned the material with $T <$ 1.5$\times$10$^9$ K).
We will discuss this point in more detail when we will present our
nucleosynthesis results.

\section{Nucleosynthesis in multi-dimensional SNIa}

The multidimensional SNIa simulations described in Section~2 employed
a minimal nuclear reaction network, {\bf directly included in the hydrodynamic code,} 
sufficient for a good approximation of the thermonuclear energy
release and the predicted gross chemical composition agrees well
with the expectations (Reinecke et al.~2002b). It consists 
of five species ($\alpha$-particles, $^{12}$C, $^{16}$O, $^{24}$Mg, and
$^{56}$Ni) and is intended to model the energy release of the thermonuclear
reactions only. No reaction rates are employed: all material behind the
flame front is instantaneously converted to a NSE of $^{56}$Ni and
$\alpha$-particles at high densities and to $^{24}$Mg at intermediate 
densities. Below $10^7$g/cm$^3$, no burning takes place.

{\bf Therefore we follow a minimal reaction network directly in the hydrodynamic 
simulations, and a much more extended network in a post-process step.}
Here we present the results of the more detailed study of the nuclear
abundances in the ejecta obtained by post-processing the output of the
four hydrodynamic models discussed above.

Since the multidimensional hydrodynamics scheme used in modeling
the explosions is an Eulerian one (i.e.~the grid does not move with the
fluid), in order to record temperature and density evolution as a
function of time (the necessary input for nucleosynthesis
calculations) we homogeneously distributed $\sim$10000 marker
particles (in 2D models) and $\sim$20000 marker particles (in 3D
models) and followed their $T$ and $\rho$ evolution. The number of
particles in the simulation was chosen in order to reproduce in the
best way the resolution of the grid (see discussion below).  We then
calculated the nucleosynthesis experienced by each marker and computed
the total yield as a sum over all the markers, after the decay of
unstable isotopes.

\subsection{Tracer particles method}

In one spatial dimension it is nowadays possible to solve reaction
networks consisting of hundreds of species online with the
hydrodynamics (see e.g. Rauscher et al.~2002 for explosive
nucleosynthesis calculations in core collapse SNe). However, it is
more common to use reduced networks in order to obtain the
(approximate) energy generation rate for the hydrodynamics and to  
calculate the detailed chemical composition only afterwards in a
post-processing step. This is facilitated by the lagrangian nature of
nearly all 1D codes employed for explosive nucleosynthesis
calculations.  In lagrangian schemes, the grid moves with the fluid
and therefore it is possible to record the evolution of the
temperature and density for different fluid elements (i.e. lagrangian
zones), which is required for the post-processing.  In contrast, most
grid-based multidimensional hydro schemes are of Eulerian type
(i.e. the grid is fixed in space).  To obtain the necessary data for
the post-processing we added a ``lagrangian component'' to our
Eulerian scheme in the form of marker particles that we passively   
advect with the flow in the course of the Eulerian calculation,
recording their $T$ and $\rho$ history by interpolating the
corresponding quantities on the underlying Eulerian grid.

In the 3D simulations the star is subdivided into 27$^3$ grid cells
equidistant in the integrated mass $M$(r), azimuthal angle
$\varphi$ and cos $\theta$, so that each grid cell contains the same mass.
A tracer particle was placed randomly in each of those grid cells, therefore
the total number of tracer particles we used is 27$^3 =$ 19683.
After numerical inversion of the function $M$(r), the
($M$,$\varphi$,cos $\theta$) coordinates are mapped onto the Cartesian
grid. For the 2D simulation, 100$^2$ particles are distributed in $r$
and cos $\theta$ directions, using the same procedure as above. In all
cases the simulation covers one octant, therefore to get the total
white dwarf mass we mutiply the mass of each marker by 8 and we sum  
over all the markers. They are distributed in the way to have identical 
mass, that is therefore calculated as the ratio between $M_{ch}$ and the 
total number of tracer particles. The mass of each marker is therefore between 
$\sim$10$^{-3} M_\odot$ and $\sim$10$^{-4} M_\odot$. The initial distribution 
of the markers for the {\it b30\_3d\_768} case is shown in Fig.~4.

   \begin{figure}
   \centering
    \includegraphics[width=8cm]{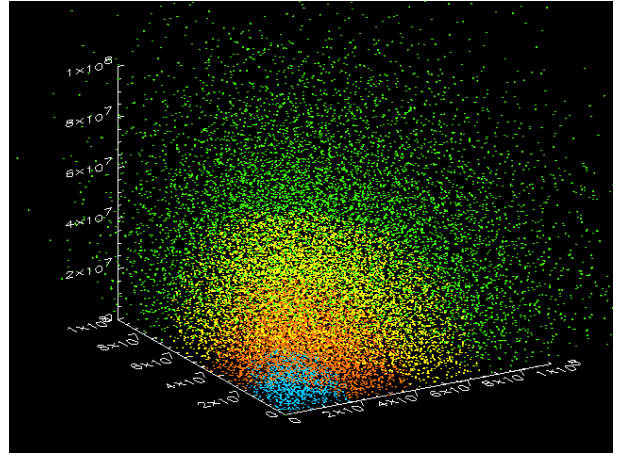}
      \caption{Radial distribution of the tracer particles in the 3D model at the   
               beginning of the simulation. \label{fig4}}
   \end{figure}

Finally, we compare the $T$(t) values for each marker extracted from 
the hydrodynamical model with the $T$(t) derived from using the
internal energy from the hydro-code (which has contributions from the
Boltzmann gas of ionized nuclei, the Planck-spectrum of photons, and
relativistic degenerate or non-degenerate electrons positrons). This
is done by calculating separately the equation of state for a given 
maker particle's internal energy, density and composition, and 
deriving from that the temperature T. As chemical composition for this
calculation we use a 16 isotopes network, composed by n, p, $^4$He,
$^{12}$C, $^{16}$O, $^{20}$Ne, $^{24}$Mg, $^{28}$Si, $^{32}$S,
$^{36}$Ar, $^{40}$Ca, $^{44}$Ti, $^{48}$Cr, $^{52}$Fe, $^{56}$Ni,  
$^{60}$Zn.  We found that the $T$(t) obtained directly from the
hydrodynamic model is in average lower by 10\% up to 20\% for markers
with high temperatures ($T \ge$ 6$\times$10$^9$ K) as compared with the
temperature derived from the energy. This can be understood from the
fact that in the hydro-code the internal energy density is the directly  
computed variable and, therefore, is more accuratly determined than
the temperature. Consequently the more precise $T$(t) distribution is 
derived from the internal energy and the actual composition. This is 
what we used for our nucleosynthesis calculations.

\subsection{Nucleosynthesis network}

The nuclear reaction network employed in computing our post-process
explosive nucleosynthesis calculations contains 383 nuclear species
ranging from neutrons, protons, and $\alpha$-particles to molibdenum.
A detailed description of the code we used to solve the nuclear
network and the reaction rate library utilized is given by Thielemann
et al.~(1996) and Iwamoto et al.~(1999).  Weak interaction rates
applied in those calculations were taken from Fuller, Fowler, \& 
Newman~(1985). More recently full large-scale shell model calculations 
for electron capture and $\beta$-decays became available also for
pf-shell nuclei, i.e. the Fe-group (Langanke \& Martinez-Pinedo~2000, 
Martinez-Pinedo et al.~2000).
They have already been included in preliminar calculations by 
Brachwitz et al.~(2000) and Thielemann et al.~(2003). We also included 
the new rates in the calculations presented in this paper. As discussed 
below more in details, the nuclear reaction rates entering the thermonuclear
modeling can play an important role. While large portions of the
ejecta which experience maximum temperatures in excess of
6$\times$10$^9$ K follow nuclear statistical equilibrium (a chemical
equilibrium of all strong and electromagnetic reactions), weak
interactions occur on a longer timescale and different choice of
Fueller et al.~(1985) (as used by Iwamoto et al.~1999) or 
Langanke \& Martinez-Pinedo~(2000) (as used for this work, and 
by Brachwitz et al.~2000, Thielemann et al.~2003), can strongly 
affect the results. 

The initial WD composition we used consists of (mass
fraction) 0.475 M$_\odot$ of $^{12}$C, 0.5 M$_\odot$ of $^{16}$O, and
0.025 M$_\odot$ of $^{22}$Ne (in agreement with the W7 initial 
composition, Iwamoto et al.~1999). With this initial composition we  
typically simulate a solar metallicity SNIa.  

When the flame passes through the fuel, $^{12}$C, $^{16}$O and
$^{22}$Ne are converted to ashes with different compositions depending
on the intial $T$ and $\rho$. We stop our nucleosynthesis calculations
after $\sim$1.5 sec. When the temperature in the markers dropped at
$\sim$1.5 10$^9$ K the explosive nucleosynthesis is
almost frozen. The distribution of T and $\rho$ as a function of time is
shown in Fig.~5 for the model {\it b30\_3d\_768}.  The {\it thick  
lines} represent the upper envelope for T and $\rho$, and with the {\it
dashed lines} we plot T and $\rho$ histories in some markers taken as
examples. As one can notice from Fig.~5, the T distribution is not yet
below $\sim$1.5 10$^9$ K for all the tracer particles. At 
T $\sim$ 3 - 4 10$^9$ K (i.e. the upper values shown in Fig.~5) we still
expect some explosive C-burning products. Therefore we linearly extrapolate 
T and $\rho$ until all the particles have T $<$ 1.5 10$^9$ K 
(that corresponds to $\sim$ 1 sec.).

   \begin{figure}
   \centering
    \includegraphics[width=9cm]{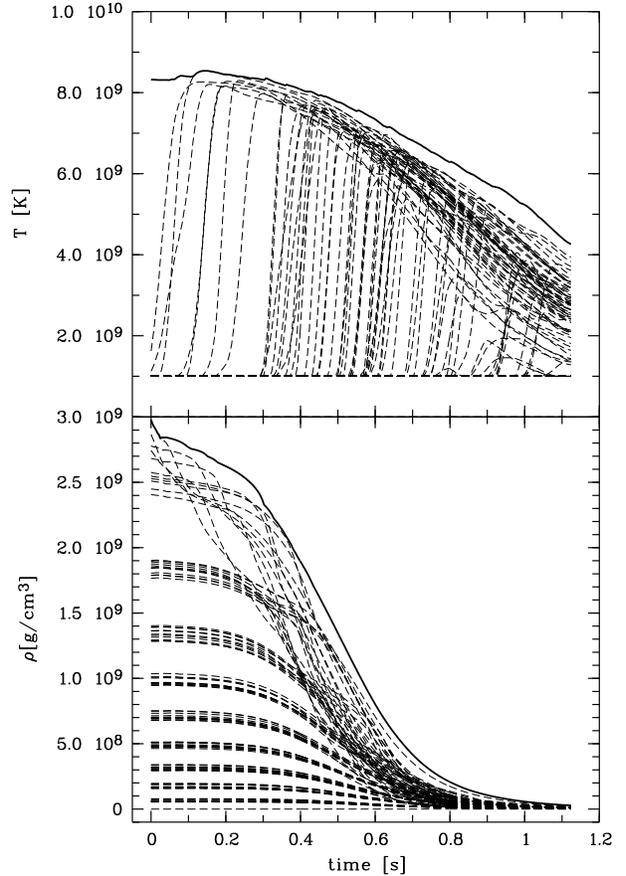}
      \caption{Temperature ({\it upper panel}) and density ({\it lower panel})
               history in the tracer particles for the {\it
               b30\_3d\_768} model. The {\it thick lines} represent
               the upper envelope of the distribution and the {\it thin dashed lines}
               represent some of the markers randomly taken as examples. \label{fig5}}
   \end{figure}

The combination of T and $\rho$ vs. time in each marker is very 
important for the nucleosynthesis results (as discussed
below). Comparing the model shown in Fig.~5 with the T and $\rho$
distribution shown by Iwamoto et al.~(1999) for their W7 model, we
note differences that can be interesting for the nucleosynthesis
calculations. First, the timescale in our models are much smaller
($\sim$1.5 sec) with respect to the W7 timescale ($\sim$6 sec). Then
different combinations of T and $\rho$, in our case rather low 
T at still high $\rho$, also give us interesting differences in the   
nucleosynthesis calculations.

\section{Discussion and results}

In this Section we present the results for nucleosynthesis
calculations in the model {\it c3\_2d\_512}, {\it c3\_3d\_256},
{\it b5\_3d\_256}, and our 'standard' {\it b30\_3d\_768}.
We also compare them with the W7 calculations by Brachwitz et
al.~(2000) and Thielemann et al.~(2003) (note that in Figures and 
Tables we better compare with Brachwitz et al. calculations instead 
of Iwamoto et al.~1999, in order to be consitent with the use of elecetron
capture rates). We analyze the consequences of different hydrodynamic
resolutions on the nucleosynthesis, we compare their different distribution of
burned and unburned material, we discuss the trend of the Y$_e$
in the markers as a consequence of our nucleosynthesis calculations. 
Finally, we discuss the velocity distribution of different nuclear species.

\subsection{Nucleosynthesis and yields: comparison between 2D and 3D}

For the nucleosynthesis calculations the peak temperatures combined
with the density distribution achieved during
the propagation of the front, are the most important quantities. As
shown in Fig.~5, for the {\it b30\_3d\_768} model the maximum of the T
in the markers covers a large range (1.5$\times$10$^9$ K $< T <$
8.4$\times$10$^9$ K). Note that in this Figure each {\it dotted line}
represents a tracer particle trend (we selected randomly some markers
for this plot), and the {\it thick line} is the upper envelope of the
$T$ and $\rho$ distribution in the total tracer particle sample.
At high temperatures all strong interactions and
photodisintegrations are so fast that a chemical equilibrium 
(nuclear statistical equilibrium, NSE) is immediately achieved (in our
calculations we assume NSE condition for $T \ge 6.0\times$10$^9$K).
The resulting chemical composition is therefore only dependent on
$\rho$, $T$, and the neutron eccess (that is determined by the total  
amount of electron captures taking place on longer timescales).  An   
example of the behaviour of the chemical abundances in one marker as a
function of time is shown in Fig.~6. For this case the marker is
originally located at a radius of $\sim$150 km, i.e.  in the innermost 
dense zone; the initial density at the position of the marker is
$\sim$ 2.5$\times$10$^9$ gr/cm$^3$ (the central density for this     
model is 2.9$\times$10$^9$ gr/cm$^3$) and it is heated by the flame 
front almost immediately (at
$\sim$0.1 s). The NSE conditions are achieved very fast and the
temperature reaches a quite high peak $T\sim$8.5$\times$10$^9$K
(Fig.~6, upper left panel). The resulting electron fraction $Y_e$  
drops rapidly (in $\sim$0.3 s) from the initial value of 0.4989 to
0.468 (Fig.~6, upper right panel).  The main O-burning products,   
$^{28}$Si and $^{32}$S are shown in Fig.~6 (left middle panel),
together with the abundance of $^{24}$Mg resulting from
C-burning. Furthermore, the rise of the temperature in excess of
6$\times$10$^9$ K leads to a complete NSE and $^{28}$Si exhaustion   
within 0.1 s. Due to the high density (typical of explosive
Si-burning) normal freeze-out occurs. One of the main products is 
$^{56}$Fe (Fig.~6, right middle panel).  Since $T$ and $\rho$
conditions are very high, also neutron-rich nuclei are built up due 
to electron captures, and $^{56}$Fe is partly replaced by $^{54}$Fe 
(Fig.~6, right middle panel) and $^{58}$Ni (Fig.~6, left lower
panel). In such details we can follow the nucleosynthesis changes 
along the time in all the tracer particles.

   \begin{figure}
   \centering
    \includegraphics[width=9cm]{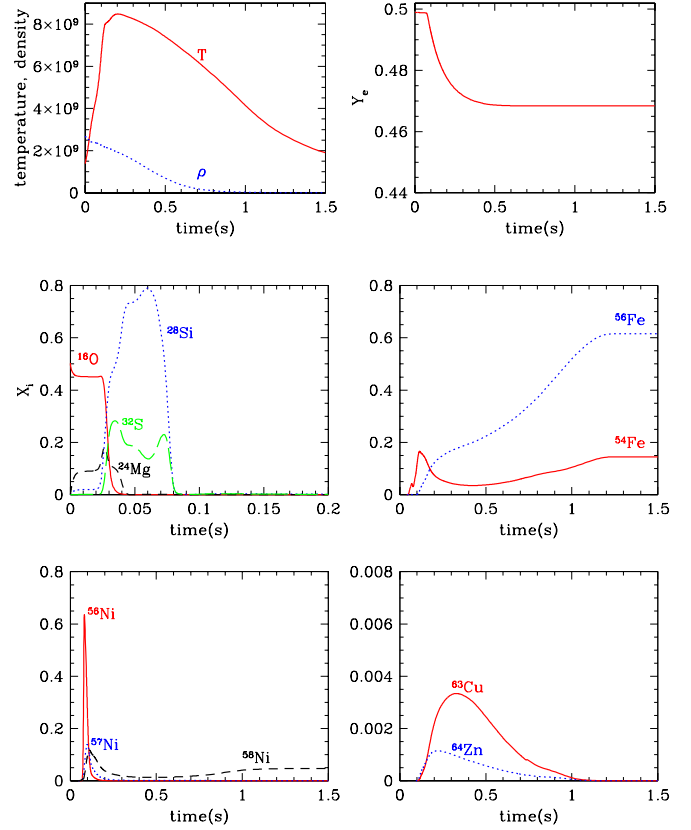}
      \caption{Example of the nucleosynthesis calculation in one tracer particle.      
               T and $\rho$ are plotted in the upper left panel; with 
               T $>$ 6$\times$10$^{9}$ K reaches NSE conditions. In
               the upper right panel the resulting Y$_e$ is shown. The other panels
               give the mass fraction vs. time for $^{16}$O, $^{28}$Si,
               $^{32}$S, $^{54,56}$Fe, $^{56,57,58}$Ni, $^{63}$Cu, and
               $^{64}$Zn. For $^{16}$O, $^{28}$Si, $^{32}$S
               the time is plotted only up to 0.2 s since their
               abundances are zero at later times. \label{fig6}}
   \end{figure}

\begin{table*}
{\scriptsize
\begin{center} TABLE 1\\
{\sc Synthesized Mass ($M_\odot$) in SNIa models}\\
\vspace{1.0em}
\begin{tabular}{ccccccc}
\hline\hline
 Species & W7$^{(\rm a)}$ & {\it c3\_2d\_512}$^{(\rm b)}$ & {\it c3\_3d\_256}$^{(\rm b)}$ &
{\it c3\_3d\_256}$^{(\rm c)}$ & {\it b5\_3d\_256}$^{(\rm c)}$ & {\it b30\_3d\_768}$^{(\rm
c)}$ \\ \hline

$^{12}$C  & 5.04E-02 & 4.09E-01 & 3.37E-01 & 3.37E-01 & 2.49E-01 & 2.78E-01 \\
$^{13}$C  & 1.07E-06 & 1.07E-10 & 9.71E-11 & 9.68E-11 & 8.21E-06 & 3.98E-06 \\
$^{14}$N  & 4.94E-07 & 2.71E-09 & 3.96E-09 & 3.48E-09 & 1.04E-03 & 2.76E-04 \\
$^{15}$N  & 1.25E-09 & 4.40E-11 & 7.18E-11 & 6.99E-11 & 2.48E-05 & 1.23E-06 \\
$^{16}$O  & 1.40E-01 & 4.74E-01 & 4.16E-01 & 4.17E-01 & 3.90E-01 & 3.39E-01 \\
$^{17}$O  & 3.05E-08 & 1.16E-09 & 1.29E-09 & 1.13E-09 & 7.81E-06 & 1.31E-06 \\
$^{18}$O  & 7.25E-10 & 9.49E-11 & 1.62E-10 & 1.52E-10 & 1.15E-04 & 1.01E-05 \\
$^{19}$F  & 5.72E-10 & 2.64E-11 & 3.72E-11 & 3.34E-11 & 1.08E-06 & 2.84E-08 \\
$^{20}$Ne & 1.97E-03 & 4.70E-03 & 7.39E-03 & 7.10E-03 & 3.18E-02 & 6.28E-03 \\
$^{21}$Ne & 8.51E-06 & 7.11E-07 & 1.14E-06 & 1.03E-06 & 5.96E-05 & 2.16E-05 \\
$^{22}$Ne & 2.27E-03 & 2.15E-02 & 1.77E-02 & 1.77E-02 & 1.14E-02 & 1.42E-02 \\
$^{23}$Na & 6.20E-05 & 2.99E-05 & 5.09E-05 & 5.10E-05 & 3.49E-03 & 8.65E-04 \\
$^{24}$Mg & 1.31E-02 & 1.04E-02 & 1.48E-02 & 1.26E-02 & 2.35E-02 & 7.53E-03 \\
$^{25}$Mg & 4.71E-05 & 5.49E-05 & 8.57E-05 & 7.64E-05 & 2.41E-03 & 5.13E-04 \\
$^{26}$Mg & 3.31E-05 & 6.60E-05 & 1.06E-04 & 1.01E-04 & 8.56E-04 & 1.81E-04 \\
$^{27}$Al & 8.17E-04 & 7.39E-04 & 1.08E-03 & 9.73E-04 & 2.11E-03 & 5.85E-04 \\
$^{28}$Si & 1.52E-01 & 4.42E-02 & 5.89E-02 & 5.39E-02 & 1.19E-01 & 5.39E-02 \\
$^{29}$Si & 7.97E-04 & 6.47E-04 & 9.49E-04 & 9.22E-04 & 1.81E-03 & 5.61E-04 \\
$^{30}$Si & 1.43E-03 & 1.06E-03 & 1.48E-03 & 1.31E-03 & 2.20E-03 & 8.03E-04 \\
$^{31}$P  & 3.15E-04 & 2.02E-04 & 2.85E-04 & 2.69E-04 & 5.24E-04 & 1.72E-04 \\
$^{32}$S  & 8.45E-02 & 1.60E-02 & 2.22E-02 & 2.57E-02 & 5.70E-02 & 2.62E-02 \\
$^{33}$S  & 4.11E-04 & 1.05E-04 & 1.42E-04 & 1.58E-04 & 3.21E-04 & 1.21E-04 \\
$^{34}$S  & 1.72E-03 & 8.68E-04 & 1.15E-03 & 1.15E-03 & 2.30E-03 & 1.04E-03 \\
$^{36}$S  & 2.86E-07 & 1.64E-07 & 2.24E-07 & 2.47E-07 & 3.95E-07 & 1.53E-07 \\
$^{35}$Cl & 1.26E-04 & 3.60E-05 & 4.88E-05 & 5.90E-05 & 1.31E-04 & 4.58E-05 \\
$^{37}$Cl & 3.61E-05 & 6.89E-06 & 8.97E-06 & 1.27E-05 & 3.21E-05 & 1.21E-05 \\
$^{36}$Ar & 1.49E-02 & 2.12E-03 & 3.14E-03 & 4.09E-03 & 9.04E-03 & 4.24E-03 \\
$^{38}$Ar & 8.37E-04 & 3.30E-04 & 4.13E-04 & 5.12E-04 & 1.20E-03 & 5.59E-04 \\
$^{40}$Ar & 1.38E-08 & 1.49E-09 & 2.06E-09 & 3.04E-09 & 4.92E-09 & 1.91E-09 \\
$^{39}$K  & 6.81E-05 & 1.51E-05 & 1.84E-05 & 2.95E-05 & 7.69E-05 & 3.24E-05 \\
$^{41}$K  & 6.03E-06 & 9.03E-07 & 1.17E-06 & 2.20E-06 & 6.03E-06 & 2.41E-06 \\
$^{40}$Ca & 1.21E-02 & 1.68E-03 & 2.66E-03 & 3.40E-03 & 7.08E-03 & 3.59E-03 \\
$^{42}$Ca & 2.48E-05 & 6.66E-06 & 8.43E-06 & 1.41E-05 & 3.61E-05 & 1.58E-05 \\
$^{43}$Ca & 1.07E-07 & 2.26E-08 & 3.06E-08 & 3.96E-08 & 6.37E-08 & 5.10E-08 \\
$^{44}$Ca & 9.62E-06 & 1.80E-06 & 2.81E-06 & 3.10E-06 & 4.52E-06 & 3.61E-06 \\
$^{46}$Ca & 2.44E-09 & 2.58E-12 & 3.46E-12 & 1.14E-11 & 1.91E-11 & 8.53E-12 \\
$^{48}$Ca & 1.21E-12 & 1.99E-17 & 3.20E-17 & 1.05E-16 & 1.54E-16 & 4.01E-15 \\
$^{45}$Sc & 2.17E-07 & 2.16E-08 & 3.06E-08 & 6.08E-08 & 1.65E-07 & 6.47E-08 \\
$^{46}$Ti & 1.16E-05 & 2.80E-06 & 3.53E-06 & 5.62E-06 & 1.47E-05 & 6.62E-06 \\
$^{47}$Ti & 5.45E-07 & 1.38E-07 & 1.88E-07 & 2.20E-07 & 3.61E-07 & 2.64E-07 \\
$^{48}$Ti & 2.07E-04 & 4.11E-05 & 6.96E-05 & 7.28E-05 & 1.32E-04 & 7.69E-05 \\
$^{49}$Ti & 1.59E-05 & 3.28E-06 & 5.22E-06 & 5.62E-06 & 1.10E-05 & 5.78E-06 \\
$^{50}$Ti & 1.62E-06 & 8.22E-10 & 2.08E-08 & 2.08E-08 & 8.67E-10 & 2.67E-07 \\
$^{50}$V  & 4.58E-09 & 2.04E-09 & 3.66E-09 & 3.50E-09 & 2.69E-08 & 2.66E-09 \\
$^{51}$V  & 3.95E-05 & 1.70E-05 & 1.90E-05 & 1.90E-05 & 2.89E-05 & 1.95E-05 \\
$^{50}$Cr & 2.23E-04 & 1.20E-04 & 1.20E-04 & 1.10E-04 & 1.67E-04 & 1.19E-04 \\
$^{52}$Cr & 4.52E-03 & 1.91E-03 & 2.80E-03 & 2.76E-03 & 3.48E-03 & 2.58E-03 \\
$^{53}$Cr & 6.49E-04 & 4.78E-04 & 5.18E-04 & 4.81E-04 & 5.09E-04 & 4.83E-04 \\
$^{54}$Cr & 3.04E-05 & 3.42E-06 & 6.38E-06 & 5.92E-06 & 4.11E-06 & 1.22E-05 \\
$^{55}$Mn & 6.54E-03 & 5.63E-03 & 5.93E-03 & 5.53E-03 & 4.53E-03 & 6.38E-03 \\
$^{54}$Fe & 7.49E-02 & 6.79E-02 & 6.61E-02 & 6.21E-02 & 4.48E-02 & 7.33E-02 \\
$^{56}$Fe & 6.69E-01 & 2.44E-01 & 3.28E-01 & 3.36E-01 & 3.40E-01 & 4.39E-01 \\
$^{57}$Fe & 2.52E-02 & 1.05E-02 & 1.35E-02 & 1.36E-02 & 1.28E-02 & 1.86E-02 \\
$^{58}$Fe & 1.74E-04 & 8.25E-06 & 3.16E-05 & 3.02E-05 & 8.58E-06 & 1.05E-04 \\
$^{59}$Co & 7.66E-04 & 6.70E-04 & 7.62E-04 & 6.81E-04 & 4.53E-04 & 7.33E-04 \\
$^{58}$Ni & 1.02E-01 & 6.13E-02 & 7.52E-02 & 7.31E-02 & 5.56E-02 & 9.66E-02 \\
$^{60}$Ni & 9.22E-03 & 7.23E-03 & 9.24E-03 & 8.16E-03 & 5.39E-03 & 7.73E-03 \\
$^{61}$Ni & 2.69E-04 & 6.11E-05 & 8.86E-05 & 9.26E-05 & 9.99E-05 & 1.13E-04 \\
$^{62}$Ni & 2.31E-03 & 5.71E-04 & 7.78E-04 & 8.16E-04 & 9.21E-04 & 1.12E-03 \\
$^{64}$Ni & 1.84E-07 & 2.73E-11 & 1.61E-09 & 1.61E-09 & 1.93E-10 & 5.29E-08 \\
$^{63}$Cu & 1.59E-06 & 9.24E-07 & 9.26E-07 & 9.27E-07 & 8.20E-07 & 9.56E-07 \\
$^{65}$Cu & 7.72E-07 & 1.88E-07 & 2.51E-07 & 2.61E-07 & 2.81E-07 & 3.77E-07 \\
$^{64}$Zn & 1.50E-05 & 3.72E-06 & 4.47E-06 & 4.65E-06 & 4.83E-06 & 6.78E-06 \\
$^{66}$Zn & 1.31E-08 & 6.11E-06 & 7.55E-06 & 7.86E-06 & 8.90E-06 & 1.16E-05 \\
$^{67}$Zn & 1.18E-11 & 4.15E-09 & 5.49E-09 & 5.68E-09 & 6.53E-09 & 7.96E-09 \\
$^{68}$Zn & 2.66E-10 & 2.85E-09 & 3.68E-09 & 3.86E-09 & 4.86E-09 & 5.26E-09 \\

\hline
\hline
\end{tabular}
\end{center}

\begin{list}{}{}

\item[$^{\mathrm{a}}$] Thielemann et al.~(2003)
\item[$^{\mathrm{b}}$] This work
\item[$^{\mathrm{c}}$] This work. In this run we allow the nucleosynthesis calculations 
for those tracer particles that reach NSE conditions only starting at 90\% of the temperature
peak.

\end{list}

}
\end{table*}

In Fig.~7 we show the distribution of Y$_e$ vs. time obtained as a 
result of the nucleosynthesis calculation in the model {\it b30\_3d\_768}. 
The two thick lines stand for the upper and lower values of the Y$_e$ in 
the markers. The dashed lines represent the Y$_e$ time evolution in some
markers randomly taken as examples. The range covers values from $\sim$0.5
(that represent the typical initial composition as well as the       
composition of markers with rather low $T$), down to $\sim$0.462,   
reached by the markers with the highest $T$ (see e.g. the example in
Fig.~6).

\begin{figure}
  \centering
   \includegraphics[angle=-90,width=9cm]{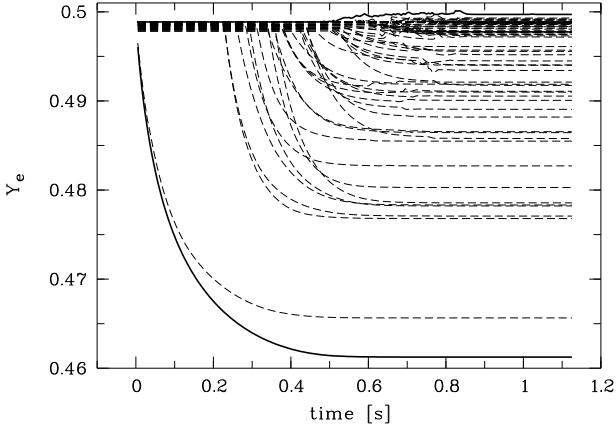}
  \caption{The same as Fig.~5 for the history of the Y$_e$ in the markers.
    \label{fig7}}
\end{figure}

Table~1 lists the synthesized masses for all the stable
isotopes up to $^{68}$Zn for the models {\it c3\_2d\_512}, {\it   
c3\_3d\_256}, {\it b5\_3d\_256}, and {\it b30\_3d\_768}. For comparison
we also include in column~2 the calculations for
the W7 model (from Thielemann et al.~2003). For the calculations presented 
in column 5,6,7 we include the nucleosynthesis results starting only when 
the temperature has reached 90\% of the peak. In fact, due to finite      
numerical resolution in the hydrodynamic simulation, the rise of the
temperature is not as steep as it would be in reality. Consequently, when
markers reach NSE conditions and weak-interactions start to play the
most important role, the nucleosynthesis timescales are fast, and
even 0.1 s (i.e. typical timescale we have for the rise of T) are 
crucial for some reactions to give an important
contribution. Nevertheless, as one can see from the Table comparing
column~4 (the nucleosynthesis for the {\it c3\_3d\_256} model has been
calculated considering the all rise of the T curve) and column~5 (the 
nucleosynthesis for the same {\it c3\_3d\_256} model has been
calculated only when T has reached 90\% of the peak), the differences
in the total yields are very
small. This is due to the fact that the amount of markers affected
by this inaccuracy is a small fraction of the total. In the Table    
we do not include isotopes heavier than $^{68}$Zn, even if the network
we used was extended up to $^{98}$Mo, since their resulting mass fraction  
are smaller than $\sim$10$^{-15} M_\odot$. An important thing to notice
is the difference in the amount of unburned material (as defined at the
beginning of this paper, we consider unburned the material with $T <$
1.5$\times$10$^9$ K) in the four models. For the 2D model {\it    
c3\_2d\_512} $\sim$60\% of the total material
remains unburned, instead of 40\% we obtain for the 3D model {\it
c3\_3d\_256}. Both of these models are centrally ignited, therefore
the difference in the amount of unprocessed material is mainly a
consequence of the difference in the total energy distribution       
due to multi-dimension effects (see Fig.~1). In Fig.~8 we show
the distribution of burned and unburned material for the model 
{\it b30\_3d\_768} ({\it upper panel}), {\it c3\_3d\_256} 
({\it middle panel}), and {\it b5\_3d\_256} ({\it lower panel}), in 
terms of number of tracer particles. The distribution plotted in Fig.~8 
has been taken at 1.1 sec for all the three models, that is the final 
time we reached in the {\it b30\_3d\_768}. For all these three simulations 
the burned component dominates at a radius of $\sim$5.0$\times$10$^7$cm, 
instead the unburned material is more or less uniformly distributed, with 
a dominant component in the outermost zones and a tail in the central regions.
As also discussed by Reinecke et al.~(2002b) the distribution of the 
unprocessed material depends on the initial conditions for the burning.
When the model is centrally ignited, as {\it c3\_3d\_256}, the dominant
component of unburned material is in the outermost regions. Instead with a
floating-bubble model with comparable resolution, like {\it b5\_3d\_256},
unprocessed material can also be concentrated in the center.
Nevertheless the amount of unprocessed material in the center for a
floating-bubble model depends on the amount of ignition-spots together with the 
resolution used. In fact (see Fig.~8) for the {\it b30\_3d\_768} case 
most of the unburned component in the innermost regions disappeared.

\begin{figure}
  \centering
  \includegraphics[width=8cm]{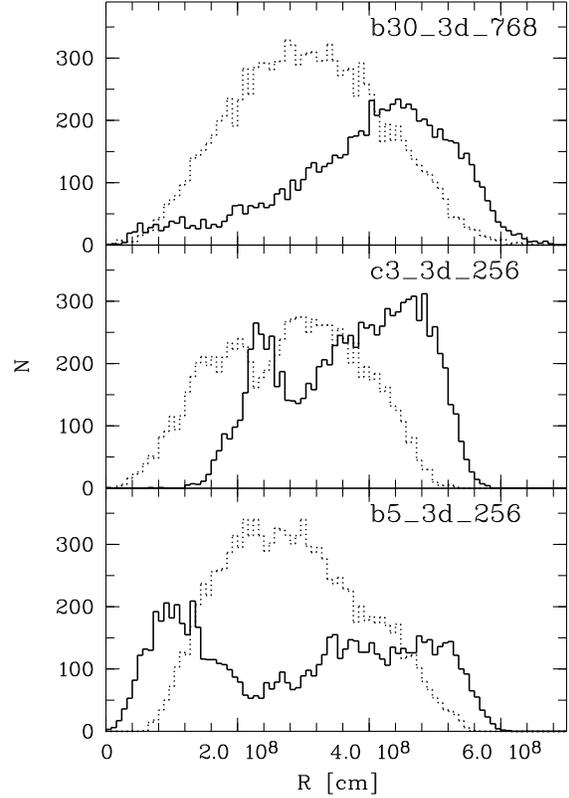}
  \caption{Distribution of the tracer particles vs. radius at $\sim$1.2 
     sec. for {\it b30\_3d\_768} ({\it upper panel}), {\it
      c3\_3d\_256} ({\it middle panel}), and {\it
      b5\_3d\_256} ({\it lower panel}) models. The unburned particles
    ($T <$ 1.5$\times$10$^9$ K) are plotted with a {\it
      solid line}, and the processed particles with a {\it
      dashed line}. 
    \label{fig8}}
\end{figure}

\begin{figure*}
\centering
\includegraphics[width=10cm,angle=-90]{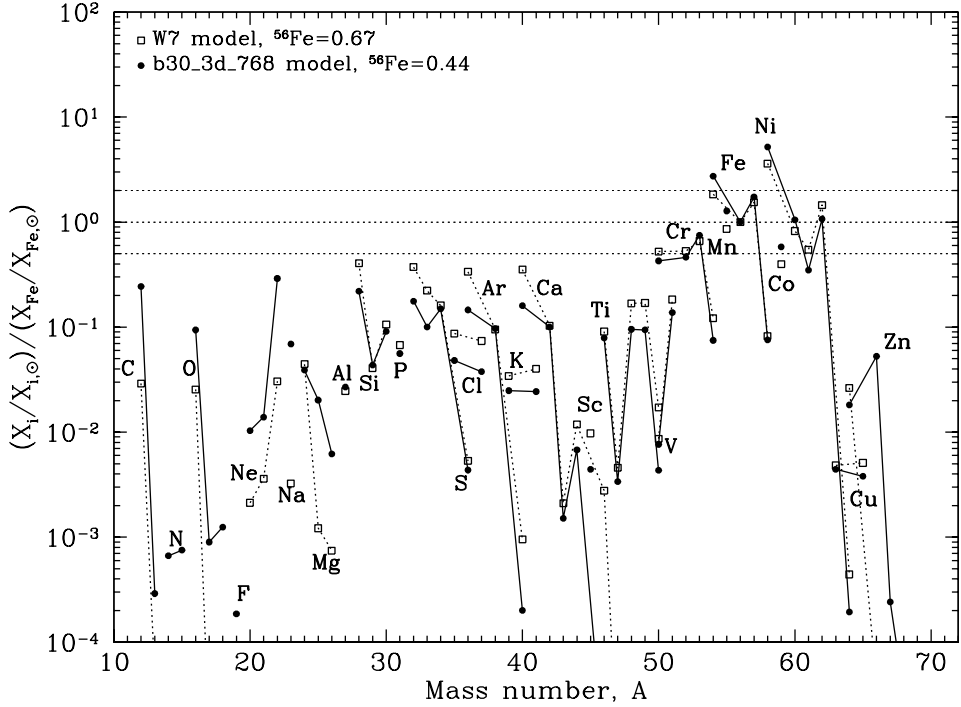}
  \caption{Nucleosynthetic yields (in mass fraction normalized to the solar
    value and to the corresponding solar ratio) obtained using 19683
    tracer particles in
    the 3D model {\it b30\_3d\_768} compared to the W7
    yields given by Thielemann et al.~(2003).
    \label{fig9}}     
\end{figure*}

We also tested consequences of burning for longer times. As mentioned in 
Section~3.2, at $\sim$1.2 sec the upper limit of the temperature is
$\sim$4$\times$10$^9$ K (see also Fig.~5), and explosive C- and a bit of 
Ne-burning can still occur. Therefore, to calculate the nucleosynthesis we 
extrapolated temperature and density for $\sim$1 sec. further (when the upper
values for $T$ are not higher that 1.5$\times$10$^9$ K). If we compare the 
nucleosynthetic yields calculated at the end of the hydro simulation with the
nucleosynthetic yields calculated using the extrapolated values of $T$ and $\rho$,
we obtain that only few \% of $^{12}$C burns. The consequences can be relevant only
for few isotopes, the main products of explosive C-burning, i.e. $^{20,21}$Ne, 
$^{23}$Na, $^{25,26}$Mg, $^{27}$Al. Therefore only for those isotopes we expect 
important changes if we could follow the hydrodynamic simulation for longer 
timescales.

It is interesting to notice the difference in the $^{12}$C and $^{16}$O   
abundance of the W7 model and our multi-dimensional models. The $^{12}$C of the
W7 is about a factor of $\sim$10 lower than in our cases, instead $^{16}$O of W7 
is only a factor of $\sim$3 lower. While $^{16}$O and $^{12}$C we obtain
are built up by unprocessed particles, in the W7 C-burning is more efficient
and burns a significant amount of $^{12}$C at low $T$ and $\rho$, with a
resulting different C/O ratio.
Also, the $^{56}$Fe mass (mainly deriving from the decay of the long-lived
$^{56}$Ni) obtained in the W7 model (0.696 $M_\odot$, Thielemann et al.~2003) 
is by far higher than the $^{56}$Fe mass resulting from the multi-dimensional 
SNIa models. The highest value we can reach in our models is obtained with the 
highest resolution floating-bubble model {\it b30\_3d\_768} (0.44 $M_\odot$).
We note that the initial conditions (in this case 30 ignition spots)
are crucial for a more precise study of the nucleosynthesis, in
particular of the innermost regions. Possibly, a model with even 
higher number of ignition bubbles would give us a still higher  
$^{56}$Fe mass.

We note that neutron-rich isotopes like $^{48}$Ca, $^{50}$Ti,
$^{54}$Cr, $^{54}$Fe, $^{58}$Ni are strongly underproduced with
respect to the W7 model presented by Iwamoto et al.~(1999), instead
are in good agreement with the Branchwitz et al.~(2000) and Thielemann 
et al.~(2003) predictions.
This is due to the differences in the electron capture rates adopted
(as just discussed above). As a consequence, electron
capture rates of nuclei affects directly the electron fraction $Y_e$
(Thielemann, Nomoto, \& Yokoi~1986). As shown in Fig.~7, the lowest 
Y$_e$ reached in the markers is 0.462, instead the lowest value in 
the W7 model by Iwamoto et al.~(1999) is $\sim$ 0.446. For the model
by Iwamoto et al.~(1999) 
low Y$_e$ values ($<$ 0.46) are reached in the innermost zones, i.e.  
$M < 0.03 M_\odot$, where the highest temperatures are reached
($\sim$9.0$\times$10$^9$ K). The highest T reached in our models, as
shown in Fig.~5, are $\sim$8.4$\times$10$^9$ K with consequently
higher Y$_e$. 

Finally, in Table~2 we report the synthesized masses of the main
radioactive species from $^{22}$Na up to $^{63}$Ni. Bigger differences
between our models and W7 are the abundances of $^{48}$Ca and
$^{66}$Zn (in this case we use for comparison the W7 data from iwamoto et 
al.~1999, since the Thielemann et al.~(2003) data are not yet available
for the long-lived isotopes). As discussed in Thielemann et al.~(2003), these 
isotopes are very sensitive to small variations in the central density 
of the model. We used 2.9$\times$10$^9$ gr/cm$^3$ instead of the 
2.0$\times$10$^9$ gr/cm$^3$ of the model B2C20 presented by Thielemann
et al.~(2003) and used in the current paper as a comparison to our model.

\begin{table*}
\begin{center} TABLE 2\\
{\sc Synthesized Mass ($M_\odot$) for radioactive species in SNIa models}\\   
\vspace{1.0em}
\begin{tabular}{ccccccc}
\hline\hline
 Species & W7$^{(\rm a)}$ & {\it c3\_2d\_512}$^{(\rm b)}$ & {\it c3\_3d\_256}$^{(\rm b)}$ &
{\it c3\_3d\_256}$^{(\rm c)}$ & {\it b5\_3d\_256}$^{(\rm c)}$ & {\it b30\_3d\_768}$^{(\rm
c)}$ \\ \hline

$^{22}$Na & 1.73E-08  & 5.76E-08 & 9.42E-08 & 8.94E-08 & 8.00E-07 & 1.00E-07 \\
$^{26}$Al & 4.93E-07  & 5.98E-07 & 9.82E-07 & 9.11E-07 & 4.53E-05 & 4.47E-06 \\
$^{36}$Cl & 2.58E-06  & 5.56E-07 & 7.62E-07 & 9.73E-07 & 1.99E-06 & 6.32E-07 \\
$^{39}$Ar & 1.20E-08  & 1.68E-09 & 2.34E-09 & 3.24E-09 & 9.41E-09 & 2.24E-09 \\
$^{40}$K  & 8.44E-08  & 7.77E-09 & 1.10E-08 & 1.73E-08 & 4.10E-08 & 1.23E-08 \\
$^{41}$Ca & 6.09E-06  & 9.01E-07 & 1.17E-06 & 2.20E-06 & 6.02E-06 & 2.40E-06 \\
$^{44}$Ti & 7.94E-06  & 1.80E-06 & 2.80E-06 & 3.09E-06 & 4.50E-06 & 3.61E-06 \\
$^{48}$V  & 4.95E-08  & 4.10E-05 & 6.95E-05 & 7.27E-05 & 1.32E-04 & 7.68E-05 \\
$^{49}$V  & 1.52E-07  & 3.28E-06 & 5.22E-06 & 5.62E-06 & 1.10E-05 & 5.78E-06 \\
$^{53}$Mn & 2.77E-04  & 4.77E-04 & 5.16E-04 & 4.78E-04 & 5.09E-04 & 4.79E-04 \\
$^{60}$Fe & 7.52E-07  & 1.44E-14 & 6.57E-12 & 6.57E-12 & 8.54E-13 & 4.52E-10 \\
$^{56}$Co & 1.44E-04  & 9.50E-05 & 1.04E-04 & 9.81E-05 & 1.69E-04 & 1.32E-04 \\
$^{57}$Co & 1.48E-03  & 1.33E-03 & 1.38E-03 & 1.16E-03 & 6.35E-04 & 1.15E-03 \\
$^{60}$Co & 4.22E-07  & 4.00E-10 & 6.82E-09 & 6.77E-09 & 1.12E-09 & 2.66E-08 \\
$^{56}$Ni & 5.86E-01  & 2.16E-01 & 2.95E-01 & 3.08E-01 & 3.27E-01 & 4.18E-01 \\
$^{57}$Ni & 2.27E-02  & 9.17E-03 & 1.21E-02 & 1.25E-02 & 1.22E-02 & 1.74E-02 \\
$^{59}$Ni & 6.71E-04  & 6.62E-04 & 7.40E-04 & 6.60E-04 & 4.47E-04 & 7.11E-04 \\
$^{63}$Ni & 8.00E-07  & 4.15E-11 & 2.10E-09 & 2.10E-09 & 1.98E-10 & 2.22E-08 \\

\hline
\hline
\end{tabular}
\end{center}

\begin{list}{}{}

\item[$^{\mathrm{a}}$] Iwamoto et al.~(1999)
\item[$^{\mathrm{b}}$] This work
\item[$^{\mathrm{c}}$] This work. In this run we allow the nucleosynthesis calculations 
for those tracer particles that reach NSE conditions only starting at 90\% of the temperature
peak.

\end{list}

\end{table*}

In Fig.~9 we show the yields obtained for our 'standard' model {\it
b30\_3d\_768} compared with the W7 yields (Brachwitz et al.~2000, Thielemann 
et al.~2003), scaled to their relative solar abundances and to the $^{56}$Fe
abundance. As one can see from the
figure, a part from the difference in the relative $^{56}$Fe mass (0.44
$M_\odot$ for the {\it b30\_3d\_768} and 0.669 $M_\odot$ for W7), as well
as in the unburned material (i.e. $^{12}$C, $^{16}$O and $^{22}$Ne),
the trend for the production of different isotopes is quite similar.

In Fig.~10 we plotted the yields of the {\it c3\_2d\_512} and
{\it c3\_3d\_256} ({\it upper panel}), and {\it b5\_3d\_256} ({\it lower panel}), 
normalized to the 'standard' model {\it b30\_3d\_768}.
From Fig.~11 one can notice that with a similar amount of burned material  
(the difference between the two models in the total burned material it
is not more than 10\%), the {\it b5\_3d\_256} produces much more
$\alpha$-elements. Infact the higher efficiency of $^{12}$C burning is
clearly shown by a higher production of $^{23}$Na and $^{40}$Ca. Also the
{\it b5\_3d\_256} model has a lower $^{56}$Fe production (0.34 $M_\odot$,
instead of 0.44 $M_\odot$ of the {\it b30\_3d\_768}), and Fe-group elements.

\begin{figure}
  \centering
  \includegraphics[width=6.5cm,angle=-90]{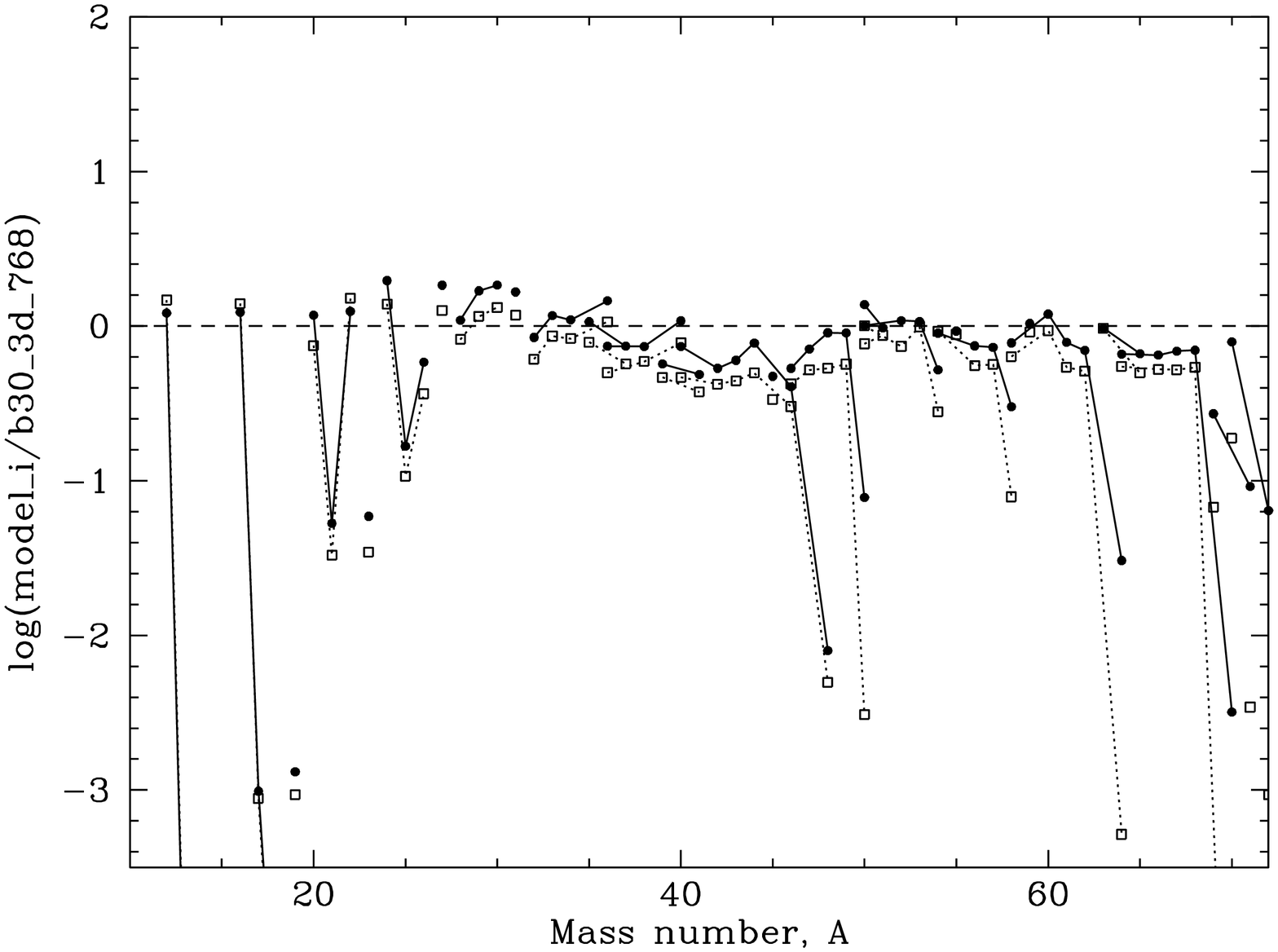}
  \includegraphics[width=6.5cm,angle=-90]{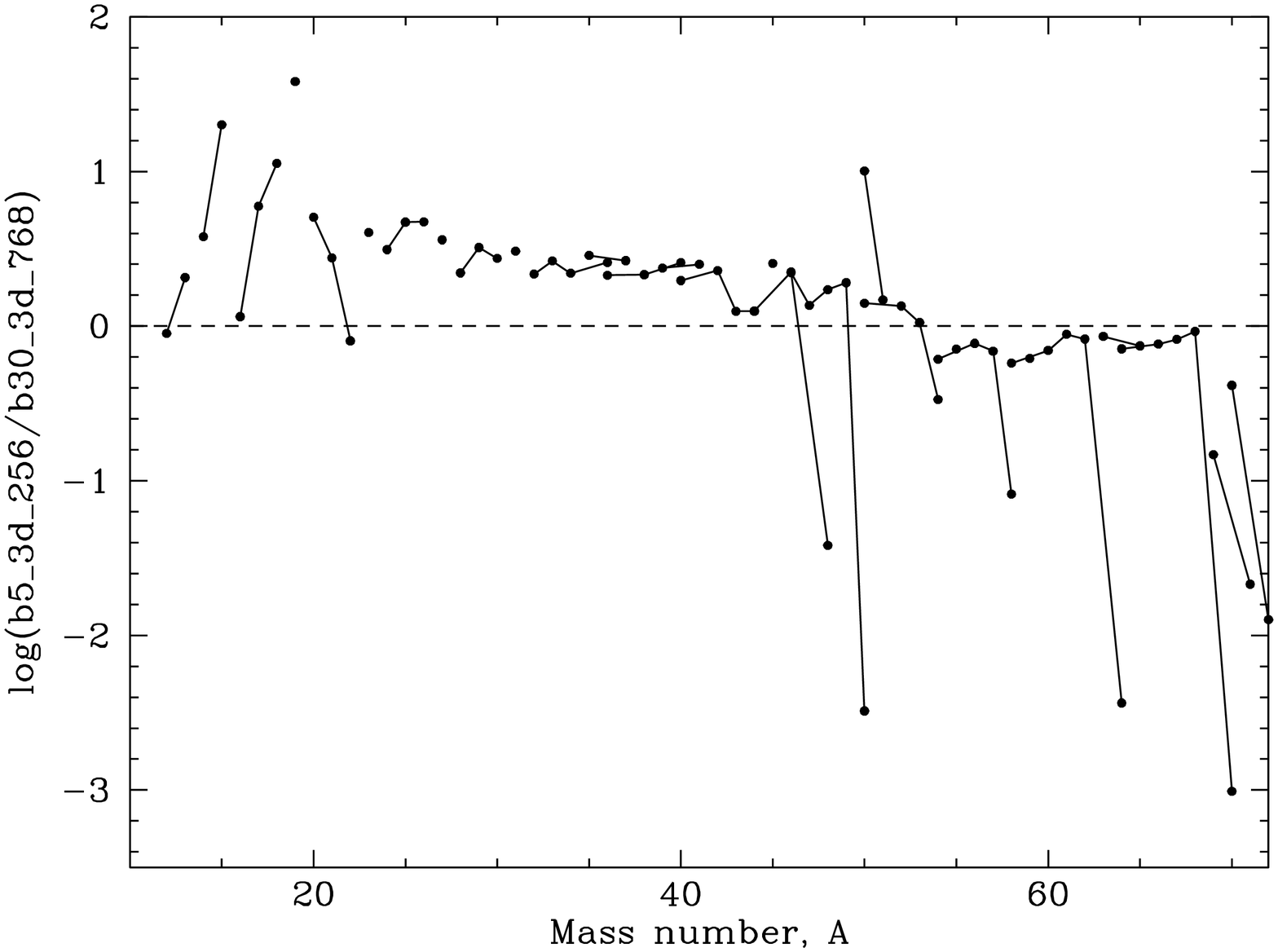}
  \caption{Nucleosynthetic yield ratio, comparing the model {\it c3\_2d\_512}
    ({\it dotted line}), and {\it c3\_3d\_256} ({\it solid
      line}), with our standard model {\it b30\_3d\_768} ({\it upper panel}).
     Nucleosynthetic yield ratio, comparing the model {\it b5\_3d\_256}
     with our standard model {\it b30\_3d\_768} ({\it lower panel}).
    \label{fig10}}
\end{figure}

\subsection{Radial velocity distribution}

In Fig.~11 we show the mass fractions of selected isotopes as a function of the radial velocity,
taken at the end of our simulation (i.e. $\sim$1.2 sec.) for the model {\it b30\_3d\_768} and
{\it c3\_3d\_256}. With {\it thick lines} we plot the unprocessed material in the form of
$^{12}$C, $^{16}$O, and $^{22}$Ne. We note that at the lowest velocities ($\sim$1000 km/s) the
dominant component is represented by the unburned material (i.e. $^{12}$C and $^{16}$O) for
the floating-bubble high-resolution model, and by $^{56}$Fe for the centrally ignited model.
On the opposite site, at the highest velocities ($>$10000 km/s) the unburned material
dominates in the centrally ignited model, instead is comparable to the $^{56}$Fe component in
the floating-bubble high resolution model. Maximum velocities reached are about 12000 km/s in 
both cases. As mentioned in the previous Section, our models at 1.2 sec are not yet in homologous 
expansion, i.e. pressure and gravity still play a role, changing the velocity distribution and 
possibly also the density, {\bf therefore the velocity distribution of the elements has to be
taken with care. However the distribution in velocitites might leave observable features 
in the spectra that could be used for diagnostic purposes. Finally, by 
projecting our 3D results on previously obtained 1D ones, the observed evolution of the Si, S, 
and Ca lines might give us the possibility to understand whether normal SNIa are well mixed 
deflagrations.}

\section{Summary and conclusions}

In this paper we presented the results of nucleosynthesis calculations obtained 
coupling a tracer particle method to two-dimensional and three-dimensional Eulerian 
hydrodynamic calculations of SNIa explosion. The multidimensional SN Ia simulations
described in this work employed a minimal nuclear reaction network, sufficient for a good
approximation of the thermonuclear energy release. Although the predicted chemical composition
agrees well with the expectations, we presented here the results of a very
detailed study of the nuclear abundances in the ejecta obtained by post-processing the 
output of the hydrodynamic models. Since the multidimensional hydrodynamical scheme applied is 
of Eulerian type (i.e. the grid does not move with the fluid), we added a lagrangian component
to the calculations in the form of tracer particles. In order to record temperature 
and density evolution as a function of time (necessary input for the nucleosynthesis 
calculations) we homogeneously distributed $\sim$20000 marker particles and followed their 
$T$  and $\rho$ evolution. We then calculated the nucleosynthesis experienced by each marker 
and computed the total yield as a sum over all the markers including the decays of unstable 
isotopes.

The nuclear reaction network employed in computing the explosive nucleosynthesis contains 
383 nuclear species ranging from neutrons, protons, and $\alpha$-particles to molibdenum. 
For this work, the initial mixture we used consists of (mass
fraction) 0.475 M$_\odot$ of $^{12}$C, 0.5 M$_\odot$ of $^{16}$O, and 0.025 M$_\odot$ of
$^{22}$Ne. When the flame passes through the fuel, C, O and Ne are converted to heavier 
elements, with different compositions depending on the $T$ and $\rho$ history. Nuclear 
statistical equilibrium conditions are assumed in the marker particles with $T >$ 6 10$^9$ K. 
At such temperatures ($T >$ 6 10$^9$ K) a mixture of $^{56}$Ni and $\alpha$-particles in NSE 
is synthesized. Below that temperature burning only produces intermediate mass elements.
Once the temperature drops $T <$ 1.5 10$^9$ K, no burning takes place during the short 
timescale ($\simeq$ 1.5 s) of the explosion (``unburned'' material).

The current research focused on the sensitivity of the explosion on the ignition 
conditions and on the detailed nucleosynthetic yields that they predict. 
We could demonstrate that multi-dimensional explosion models allow us to predict their 
nucleosynthesis yields with some confidence. It was shown that only 3D models are potentially 
able to produce enough $^{56}$Ni to explain the light curves of "normal" type Ia supernovae, 
and that also the ignition conditions (central ignition vs. several off-center ignition spot) 
affect the nucleosynthesis yields. Since the number of ignition spots we can put into the numerical 
models depend on the spatial resolution and since the explosion energy as well as the Ni-mass increase
with increasing number of spots, we expect that our best resolved {\it
  b30\_3d\_768} model is closest 
to what a "typical" pure-deflagration supernova might eject. The
general nucleosynthesis outcome of SNeIa 
is dominated by Fe-group elements, involving also sizable fractions of Si--Ca and minor amounts of 
unburned (C and O) or pure C-burning products (e.g. Na, Ne, Mg). 
Despite of the fact that differences with respect to the {\it standard} W7 nucleosynthesis
(Iwamoto et al.~1999, Brachwitz et al.~2000, Thielemann et al.~2003) are found, 
in particular in the $^{56}$Ni mass produced, as well as in the final amount of unburned 
material, in general the nuclear yields are consistent with expectations. We can therefore say 
with some confidence and without parameterization, that the Chandrasekhar mass
scenario with a pure turbulent deflagration is a viable candidate for
SN Ia explosions. We also note that the significant amount of unburned material
ejected by our SNIa models may have an interesting impact on the role of SNIa   
in the context of Galactic chemical evolution of C {\bf (giving a contribution of the 
order of $\sim$20\% to the total C at the solar composition)}. In the case of O still the 
main sources are massive stars.

Comparing the nucleosynthesis presented in this paper to observed SNIa spectra, the reader
should keep in mind that our models do not reach the homologous expansion phase. We are 
currently working to modify the combustion hydrocode, using a moving grid that will allow us 
to follow the evolution much longer. The results will be published elsewhere.
We are also performing a detailed parameter study of the variation of the central density
and of the initial carbon/oxygen ratio of the SNIa progenitor (R\"opke et al., in 
preparation). Finally, recent calculations by Timmes et al.~(2003) indicate large 
variations of the $^{56}$Ni mass as a function of metrallicity (measured by the original 
$^{22}$Ne content). An investigation of the metallicity effect on the nucleosynthesis and 
yields is also in progress (Travaglio et al., in preparation).

\begin{figure*}
  \centering
  \includegraphics[width=9cm,angle=-90]{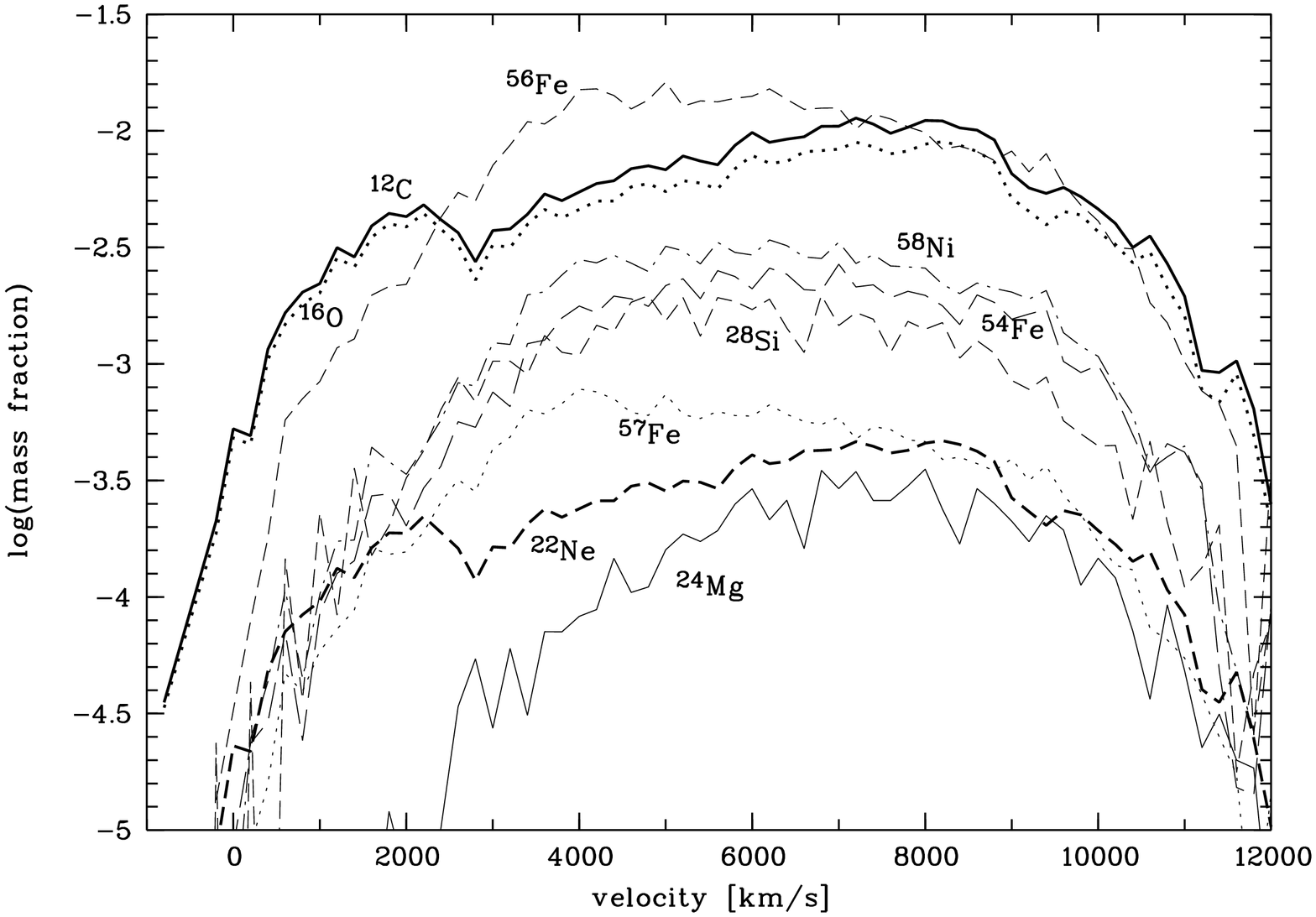}
  \includegraphics[width=9cm,angle=-90]{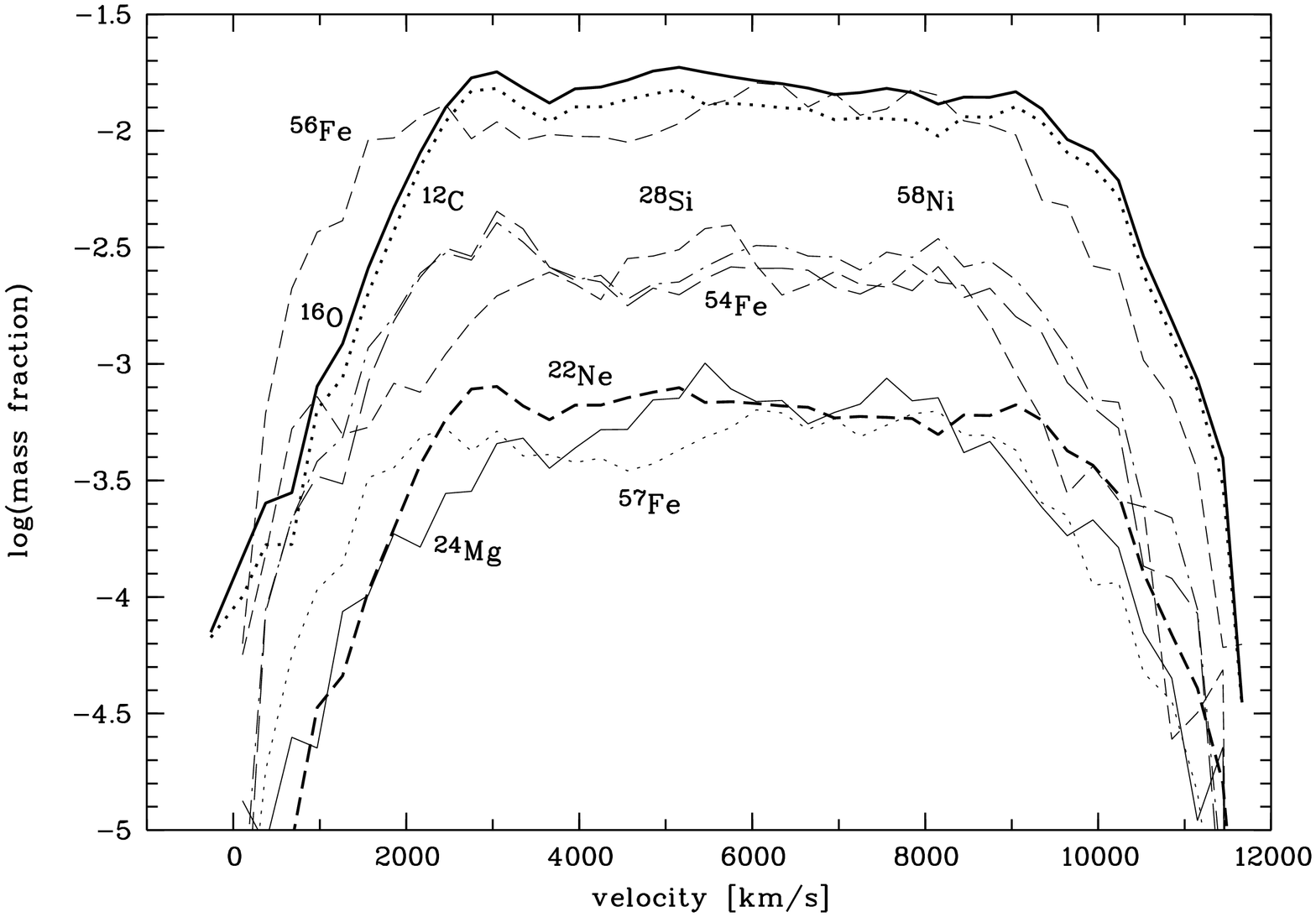}
  \caption{Mass fractions of selected isotopes as a function of the radial
    velocity of the markers (taken at $\sim$1.2 sec) for
    the {\it b30\_3d\_768} model ({\it upper panel}) and
    for the {\it c3\_3d\_256} ({\it lower panel}). The width
    of each velocity bin is 300 km/s. For each isotope we
    sum its abundance over all markers in a certain velocity bin. \label{fig11}}
\end{figure*}

{\acknowledgements C.T. thanks the Alexander von Humboldt Foundation, the
Federal Ministry of Education and Research, and the Programme for Investment
in the Future (ZIP) of the German Governmen, and the Max-Planck Institute f\"ur
Astrophysik (Garching bei M\"unchen), for their financial support.
W.H. and M.R. acknowledge support by the DFG under grant SFB-375/C5
and by the European Commission under contract HPRN-CT-2002-00303.
F.-K.T. is supported by the Swiss National Science Foundation.}

\end{document}